\documentclass[draft]{agujournal2019}
\usepackage{url} %this package should fix any errors with URLs in refs.
\usepackage{lineno}
\usepackage[inline]{trackchanges} %for better track changes. finalnew option will compile document with changes incorporated.
\usepackage{soul}

\draftfalse
\journalname{JGR: Atmospheres}

\newcommand{\HeI}{\mbox{He\textsc{i}~$\lambda10830$}}
\newcommand{\OI}{\mbox{O\textsc{i}~$\lambda8446$}}
\newcommand{\OII}{\mbox{O\textsc{i}~$\lambda11287$}}
\newcommand{\micron}{\mbox{$\mu$m}}

\begin{document}

\title{Observations of Atmospheric Helium and Oxygen with SPHEREx}

%%%%%%%%%%%%%%%%%%%%%%%%%%%%%%%%%%%%%%%%%%%%%%%
%  AUTHORS AND AFFILIATIONS
\correspondingauthor{Ryan Wills}{raw3549@rit.edu}

\authors{Howard~Hui\affil{1,2}, Chi~H.~Nguyen\affil{1}, Ryan~Wills\affil{3}, Katrina~Bossert\affil{4}, Sean~A.~Bryan\affil{4}, Kazuma~Noda\affil{3}, Jessica~Norrell\affil{4}, Yoonsoo~P.~Bach\affil{5}, James~J.~Bock\affil{1,2}, Tzu-Ching~Chang\affil{2,1}, Shuang-Shuang~Chen\affil{1}, Asantha~Cooray\affil{6}, Brendan~P.~Crill\affil{2}, Olivier~Dor\'{e}\affil{2,1}, C.~Darren~Dowell\affil{2,1}, Andreas~L.~Faisst\affil{7}, Jae~Hwan~Kang\affil{1}, Phil~M.~Korngut\affil{1,2}, Carey~M.~Lisse\affil{8}, Daniel~C.~Masters\affil{7}, Roberta~Paladini\affil{7}, Volker~Tolls\affil{9}, Michael~W.~Werner\affil{2}, Yujin~Yang\affil{5}, Michael~Zemcov\affil{3}}

\affiliation{1}{Department of Physics, California Institute of Technology, Pasadena, CA 91125, USA}
\affiliation{2}{Jet Propulsion Laboratory, California Institute of Technology, 4800 Oak Grove Drive, Pasadena, CA 91009, USA}
\affiliation{3}{School of Physics and Astronomy, Rochester Institute of Technology, 1 Lomb Memorial Dr., Rochester, NY 14623, USA}
\affiliation{4}{School of Earth and Space Exploration, Arizona State University, 781 Terrace Mall, Tempe, AZ 85287 USA}%
\affiliation{5}{Korea Astronomy and Space Science Institute (KASI), 776 Daedeok-daero, Yuseong-gu, Daejeon 34055, Republic of Korea}
\affiliation{6}{Department of Physics \& Astronomy, University of California Irvine, Irvine CA 92697, USA}%
\affiliation{7}{IPAC, California Institute of Technology, Pasadena, CA 91125, USA}
\affiliation{8}{Johns Hopkins University Applied Physics Laboratory, Laurel, MD 20723, USA}%
\affiliation{9}{Center for Astrophysics $|$ Harvard \& Smithsonian, Optical and Infrared Astronomy Division, Cambridge, MA 01238, USA}%

%%%%%%%%%%%%%%%%%%%%%%%%%%%%%%%%%%%%%%%%%%%%%%%
% KEY POINTS
%%%%%%%%%%%%%%%%%%%%%%%%%%%%%%%%%%%%%%%%%%%%%%%
\begin{keypoints}
\item SPHEREx measures near-infrared airglow from helium and oxygen across the entire sky from a 680~km low Earth orbit.
\item Helium traces the seasonal thermospheric circulation and composition changes.
\item Oxygen emission primarily responds to the geometry of the Sun illumination.
\end{keypoints}

%% \begin{abstract} starts the second page

\begin{abstract}

We present measurements of near-infrared (NIR) terrestrial airglow produced by helium and oxygen in the exosphere as observed by SPHEREx. 
Using eight months of survey data obtained from a  680~km low-Earth orbit, emission from \HeI, \OI, and \OII\ is mapped with both global spatial and multi-season temporal coverage.
These measurements are obtained along upward looking lines of sight as part of the astrophysical survey, in contrast to conventional nadir-viewing Earth remote sensing, which probes the behavior of low-density material in the thermo- and exosphere.
We describe an analytical framework to extract atmospheric emission lines in the presence of astrophysical backgrounds including stars, resolved galaxies, and the diffuse Zodiacal light.
The resulting global measurements reveal temporal variability over the survey period and systematic dependencies on geographic location.
We interpret these variations in the context of the variable Solar illumination and seasonal effects.
SPHEREx, an astrophysical space observatory, is demonstrated to be a promising new platform for monitoring NIR airglow and investigating its coupling to Solar activity and global geophysical processes.

\end{abstract}

\section*{Plain Language Summary}
% alright found the macro, just unnumbered section.
% Guideline: https://www.agu.org/publications/authors/journals/plain-language-summary
% It is listed as required for JGR atmos.

Understanding helium and oxygen in the upper atmosphere is crucial to study the near-Earth space environment.
Both gases emit light at near-infrared wavelengths through interactions with sunlight.
We present the measurements of helium and oxygen emission observed by the SPHEREx satellite over an eight-month period.
The measurements were taken at an altitude of 680~km, looking away from the Earth as part of an astrophysical survey, in contrast to conventional atmospheric studies where the satellite looks down toward the surface of the Earth.
We describe the analysis to separate the brightness of the atmospheric emission from extraterrestrial sources like stars, galaxies, and other diffuse background light in the Solar system.
The resulting measurements reveal temporal variability and systematic dependencies on the geographic location of the satellite.
We interpret these variations in the context of the variable Solar illumination and seasonal effects.
SPHEREx, an astrophysical space observatory, is demonstrated to be a promising new platform for monitoring oxygen and helium emission at high altitude, and investigating the coupling of their distribution and dynamics to Solar activity and global geophysical processes.

%%%% ---- INTRODUCTION ---- %%%%
\section{Introduction}
\label{S:intro}
Understanding helium and oxygen in the upper thermosphere is important for characterizing the composition, density structure, and dynamics of the near-Earth space environment.
Atomic oxygen is the dominant neutral species in the thermosphere and plays a central role in radiative cooling, photochemistry, and satellite surface interactions, while helium becomes increasingly important at higher altitudes due to diffusive separation \cite{Chamberlain1987,Hedin1987}.
At typical LEO altitudes between 500 and 800~km, diffusive separation produces strong altitude-dependent variations in atmospheric composition, with lighter species such as helium becoming more abundant relative to heavier constituents.
Variations in the abundance of these species trace global thermospheric circulation, seasonal transport between hemispheres, and responses to Solar and geomagnetic forcing \cite{Emmert2015}.
These composition variations influence the neutral density structure of the upper atmosphere, which affects satellite drag and orbital evolution for spacecraft in low Earth orbit \cite{Sutton2016}.
Observations of helium and oxygen emission therefore provide a diagnostic of thermospheric composition and dynamics that complements in situ measurements and empirical atmospheric models \cite{Meier1991}.

SPHEREx (Spectro-Photometer for the History of the Universe, Epoch of Reionization, and ices Explorer) is a NASA Medium-class Explorer~(MIDEX) mission launched in March 2025 that is performing an all-sky near-infrared (NIR) spectral survey designed to address a broad range of astrophysical questions through low-resolution spectroscopy \cite{Bock2025}.
Operating from a Sun-synchronous low-Earth orbit (LEO) at an altitude of approximately 680~km, SPHEREx maps the sky in 102~spectral bands spanning~0.75 to~5.0~\micron.
The spacecraft follows a polar orbit along the day/night terminator line with an orbital period of approximately 98 minutes \cite{Bryan2025}.
SPHEREx is designed with a wide field of view (FOV) and uniform sky coverage to support its core astrophysics science goals \cite{Dore2018}.
However, its operation in LEO makes the observations sensitive to diffuse foreground emission originating in the terrestrial atmosphere, as seen in SPHEREx survey exposures.
For SPHEREx, atmospheric airglow represents a foreground that must be modeled and removed in order to achieve its astrophysical science goals.

In this work, we present measurements of the NIR atmospheric airglow obtained during the first eight months of SPHEREx science survey, from 22 April through 31 December 2025.
We focus on the observational characterization and extraction of three prominent atmospheric emission features: the helium He\textsc{i} 1.083~\micron\ line (\HeI), and the oxygen O\textsc{i} emission lines near 0.845~\micron\ (\OI) and 1.128~\micron\ (\OII).
These features are robustly detected across the survey after removal of astrophysical backgrounds.
While these features have been detected in ground-based measurements \cite{Neo1969, Tinsley1976, Suzuki1983, Kaifler2022, Geach2025, Waldrop2008}, knowledge of their global spatial coverage has remained limited.
The SPHEREx survey provides a continuous data set spanning a wide range of Solar illumination geometry and terrestrial latitudes, revealing various trends of the physical drivers behind the airglow emission mechanism.

Auroral emission, driven by particle precipitation, is present in a small subset of SPHEREx observations, particularly at high geomagnetic latitudes.
The aurora observed by SPHEREx consists of multiple emission line features that are blended at the filter resolution.
Most exposures at the terrestrial poles taken during period of active aurora are not processed beyond the engineering data level due to elevated particle-induced transient rates \cite{Akeson2025}.
Although modest Solar-driven enhancements are captured by the analysis presented here, we do not include a dedicated auroral component in the spectral models.
Consequently, exposures dominated by auroral emission generally fail the model validation criteria and are excluded from the results discussed in this work.

This paper is organized as follows.
A review of the physical drivers behind the helium and oxygen airglow is given in Section~\ref{S:line_overview}.
In Section~\ref{S:reduction}, we describe the survey data products and the preparation steps to construct the one-dimensional spectra.
In Section~\ref{S:template_fitting}, the spectral modeling framework, line extraction methodology, and validation criteria are presented.
The resulting airglow brightness measurements are summarized in Section~\ref{S:result}.
The trends of the airglow are discussed in Section~\ref{S:discussion} in the context of the drivers presented in Section~\ref{S:line_overview}.
To conclude the paper, we summarize the potential of SPHEREx data for airglow science and discuss directions for future work that will build on the results presented in Section~\ref{S:conclusion}.

%%%% ---- AIRLOW OVERVIEW ---- %%%%
\section{Overview of Airglow Emission Mechanisms}
\label{S:line_overview}

%% Helium
\subsection{Metastable Helium}
\label{sS:He_overview}

\HeI\ emission originates from the metastable triplet state (He~$2\,^3\rm{S}_1$) and is produced primarily by resonant scattering of photons by metastable helium in the upper atmosphere \cite{Hunten1967}.
Early ground-based observations detected and measured the \HeI\ emission under conditions in which the upper atmosphere remained sunlit while the lower atmosphere was dark, establishing its detectability and confirming its association with the helium triplet \cite{Neo1969}.

The \HeI\ brightness depends on the Solar illumination geometry and on the column abundance of metastable helium along the line of sight.
Ground-based measurements have shown large scale spatial structure and pronounced seasonal variability, including winter hemisphere enhancements commonly referred to as the helium bulge \cite{Keating1967,Keating1968}.
Satellite-based analyses have quantified the latitudinal and seasonal structure of the \HeI\ airglow and relate these patterns to latitude dependent changes in helium abundance associated with global circulation and diffusive separation \cite{Liu2014}.
Time varying near-infrared Earth glow backgrounds have also been reported in space based observations with the \emph{Hubble Space Telescope} WFC3 IR channel, where enhanced background levels were correlated with viewing geometry and attributed to upper atmospheric emission, including contributions from the helium emission~\cite{Brammer2014}.

SPHEREx provides repeated observations over a wide range of illumination and latitude conditions that can sample the variability in \HeI\ emission.
Early evidence that the helium bulge is detectable in SPHEREx data has been reported in \citeA{Kulkarni2025}.

%% Oxygen
\subsection{Oxygen Emission and Drivers}
\label{sS:O_overview}
Another prominent feature in aurora and airglow is \OI, which arises from the radiative relaxation of the 3p \textsuperscript{3}P state to the less excited 3s \textsuperscript{3}S state.
This upper excited state can be generated directly or as a cascade from higher excited states \cite{Zipf1971}.
As such, emissions like \OII\ (3d \textsuperscript{3}D - 3p \textsuperscript{3}P) are often seen in tandem with \OI\ \cite{Erdman1987}.

There are several mechanisms by which excited states may be produced: the simplest is direct electron impact on atomic oxygen, the electrons having been excited by auroral acceleration or photoionization.
Electron impact may also cause the disassociation of O\textsubscript{2} into its excited constituents \cite{Waldrop2018}.
Another generation mechanism is recombination and involves the collision of an O\textsuperscript{+} ion with either an electron (radiative recombination) or an O\textsuperscript{-} ion (ion-ion recombination) \cite{Olson1971}.
Bowen fluorescence, the final relevant mechanism, arises from the proximity of Lyman $\beta$ (Ly$\beta$) and OI $\lambda$1027 \cite{Bowen1947, Waldrop2008, Waldrop2018}.
In this process, ground state atomic oxygen is excited by Ly$\beta$ and then cascades down to~3p~\textsuperscript{3}P following emission at 1.1287\micron, which is followed by \OI.

\begin{figure}
    \centering
    \includegraphics[width=0.45\textwidth]{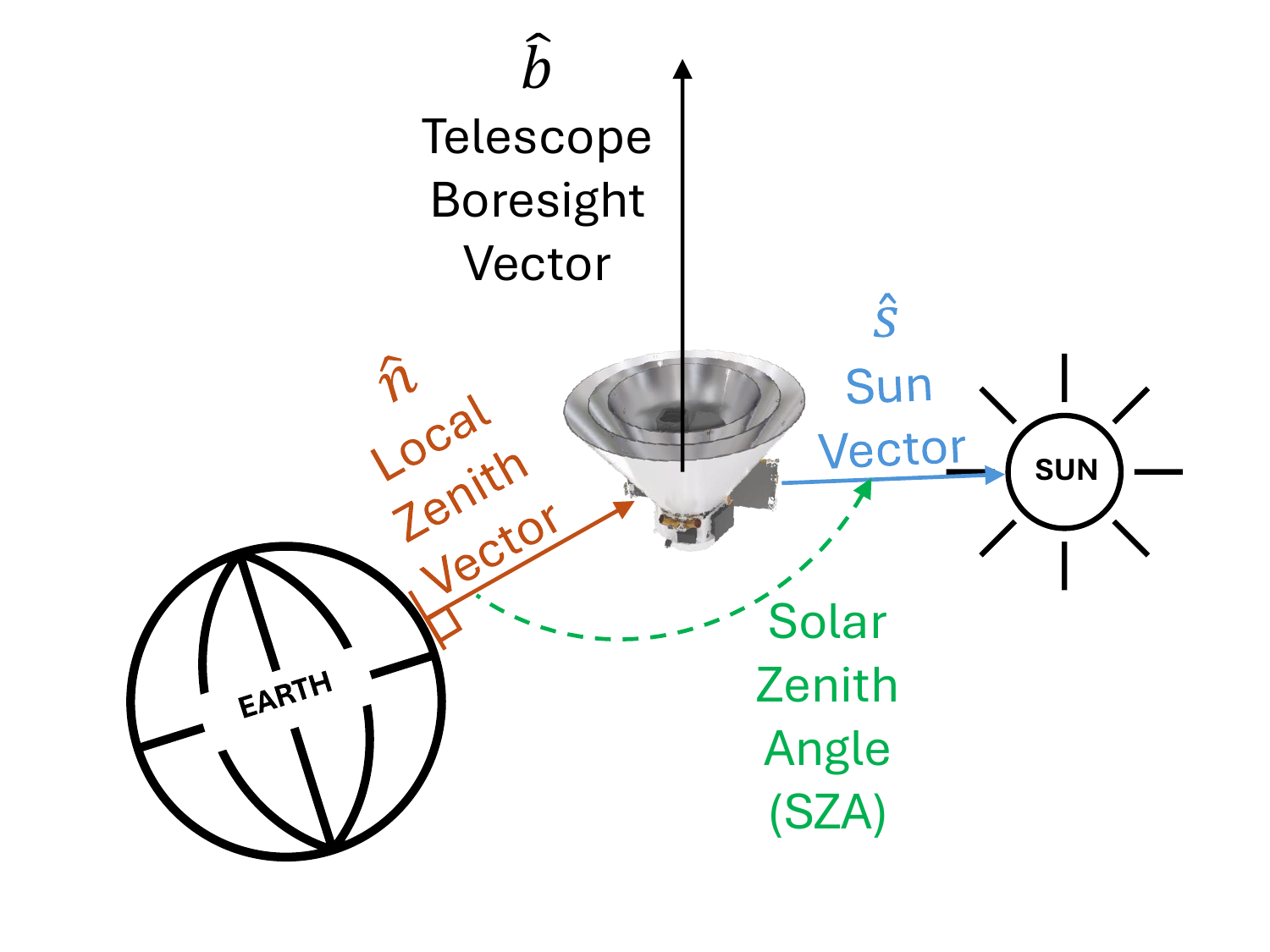}
    \caption{Illustration of the definition of the telescope boresight, zenith, and the sun vectors we use in this work.
    The solar zenith angle (green dashed line) is the signed angle between the local zenith vector and the Sun vector, taken to be positive on the sunrise side and negative on the sunset side.
    During the mission, SPHEREx has a limited allowable pointing zone \cite{Bryan2025} and only needs a limited set of vector orientations to complete the all-sky survey.
    \label{fig:SZA_schematic}}
\end{figure}

Determination of the predominant mechanism at any given time is difficult due to the multitude of sources, but several conclusions can be drawn using factors such as solar zenith angle (SZA; Figure \ref{fig:SZA_schematic}).
For example, the photoionization of electrons is confined to daylight hours and thus \OI\ emission as a result of photoelectron impact decreases as SZA increases \cite{Waldrop2008}.

In addition to Solar illumination geometry, the emission strength depends on the ambient atomic oxygen density.
Empirical atmospheric models based on satellite drag data for overall density and mass spectrometer data for relative abundances show that atomic oxygen density varies with latitude, season, and geomagnetic activity \cite{Hedin1988,Picone2002}.
Because the volume emission rate scales with both the excitation rate and the local oxygen density, spatial variations in the atmospheric composition introduce brightness variability beyond the primary SZA dependence.
Global morphology studies indicate that oxygen airglow intensity reflects the combined influence of solar illumination as well as latitude-wise and seasonal variations in the abundance of atomic oxygen \cite{Meier1991}.

%%%% ---- DATA SET AND REDUCTION ---- %%%%
\section{Survey Data and Spectral Construction}
\label{S:reduction}

%% SPHEREx Exposures
\subsection{SPHEREx Exposures and Sampling}
\label{sS:reduction_exp}

SPHEREx uses six HAWAII-2RG~(H2RG) detectors \cite{Nguyen2025} to cover the full 0.75 to 5.0~\micron\ range, labeled as Bands~1 through~6.
The instrument collects data simultaneously in six bands in 115~s integrations while continuously scanning the sky from a polar orbit along the day--night terminator.
This analysis uses calibrated science exposures from SPHEREx Bands~1 and~2, which together span wavelengths from 0.74 to 1.67~\micron\ with a nominal spectral resolving power of $R \approx 41$.

A single exposure consists of image information for $2040 \times 2040$ illuminated pixels with an angular pixel scale of $6.15$~arcseconds, corresponding to a FOV of about $3.5^{\circ} \times 3.5^{\circ}$.
Each pixel records spectrally dispersed sky brightness through a linear variable filter mounted directly above the detector, so wavelength varies monotonically along the vertical detector direction \cite{Korngut2026}.
Within this configuration, Band~1 samples 0.74 to 1.12~\micron\ and Band~2 samples 1.09 to 1.67~\micron.
Because wavelength and sky position are optically coupled, pixels at different wavelengths correspond to slightly offset sky regions, and a single exposure does not constitute a monochromatic sky map.

\begin{figure*}
    \centering
    \includegraphics[width=0.85\textwidth]{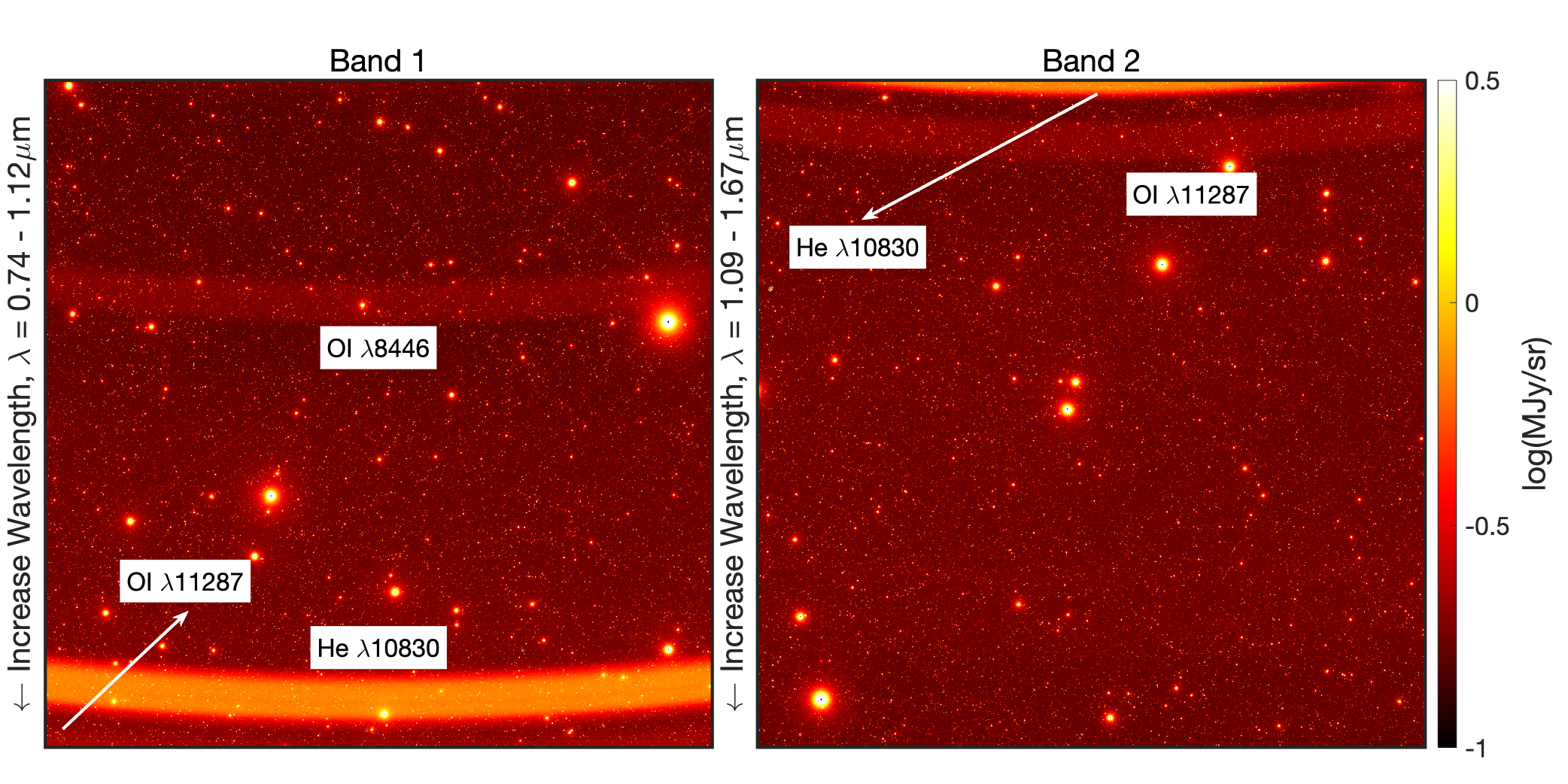}
    \caption{Example SPHEREx exposures in Band~1 and Band~2, covering wavelength ranges of 0.74 to 1.12~\micron\ and 1.09 to 1.67~\micron, respectively.
    Each spectral image spans approximately $3.5^\circ \times 3.5^\circ$ on the sky, with the wavelengths increasing from top to bottom.
    The `smile' shape is the iso-wavelength contours in the filter design \cite{Hui2026}.
    Because atmospheric airglow varies on angular scales much larger than the SPHEREx FOV at the orbital altitude, the emission appears as curved horizontal stripes across the exposure.
    \OI\ appears near the middle of Band~1 exposure.
    \HeI\ is seen near the bottom of Band~1 exposure and at the top of Band~2 exposure.
    \OII\ is found below \HeI\ in Band~2 exposure, as well as the very bottom corners in Band~1.
    }
    \label{fig:example_exposure}
\end{figure*}

The airglow emission shows little structure along the iso-wavelength direction within a single exposure.
In Figure~\ref{fig:example_exposure}, the helium and oxygen airglow features appear as horizontally extended structures across the detector in representative Band 1 and Band 2 exposures.
The extended emission suggests that the emitting layers are smooth on angular scales larger than the SPHEREx FOV at an orbital altitude of 680~km.

\begin{figure}
\centering
\includegraphics[width=0.45\textwidth]{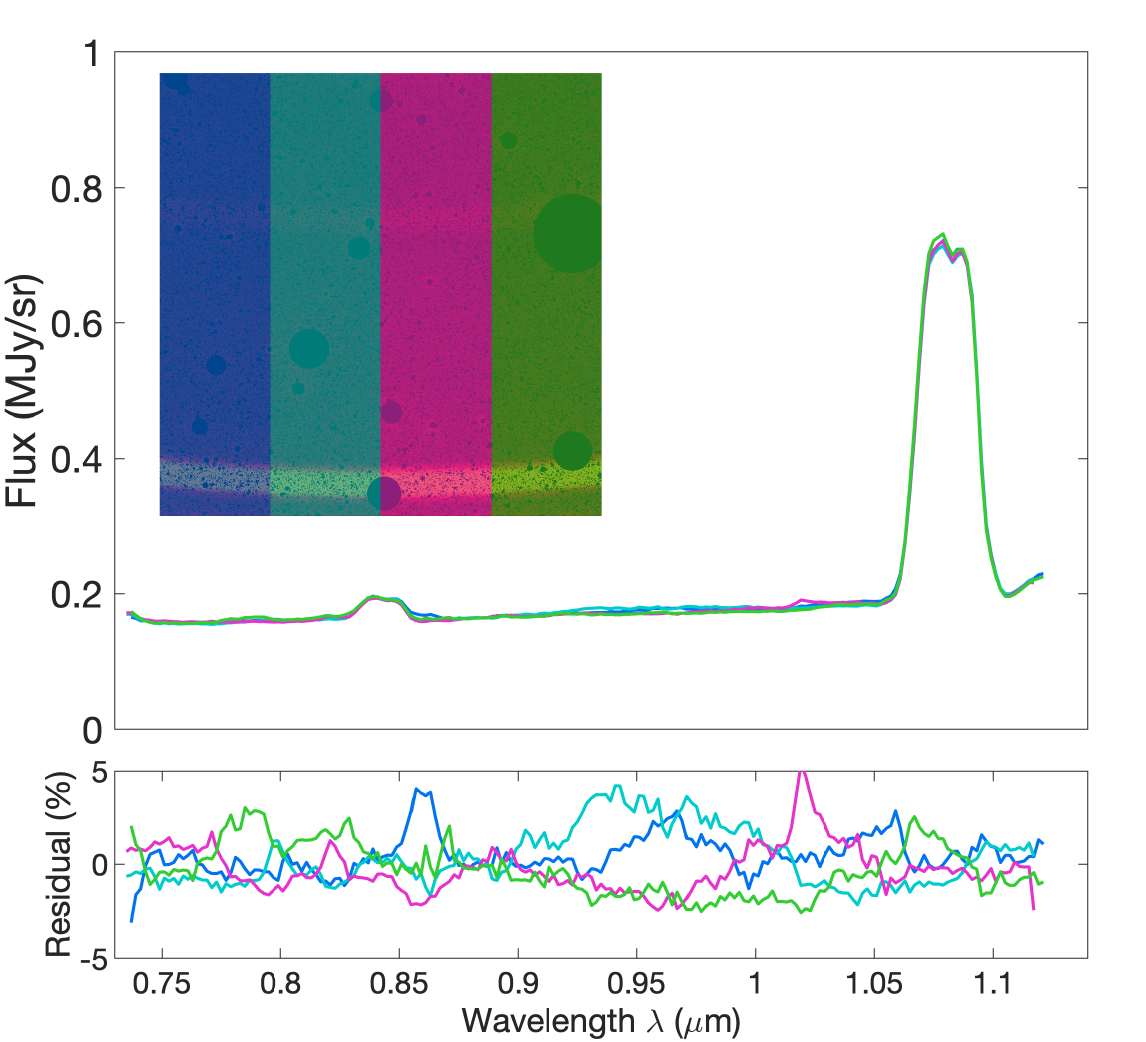}
\caption{
Example spectra illustrating the spatial uniformity of the airglow within SPHEREx FOV in a single exposure.
Top: one-dimensional spectra extracted from four regions of the exposure.
Bottom: Fractional residuals relative to the full field median.
Over most of the wavelength range, the signal is dominated by the Zodiacal light continuum, so the residual reflects the uniformity of the continuum background.
}
\label{fig:example_1D_split}
\end{figure}

\begin{figure}
\centering
\includegraphics[width=0.45\textwidth]{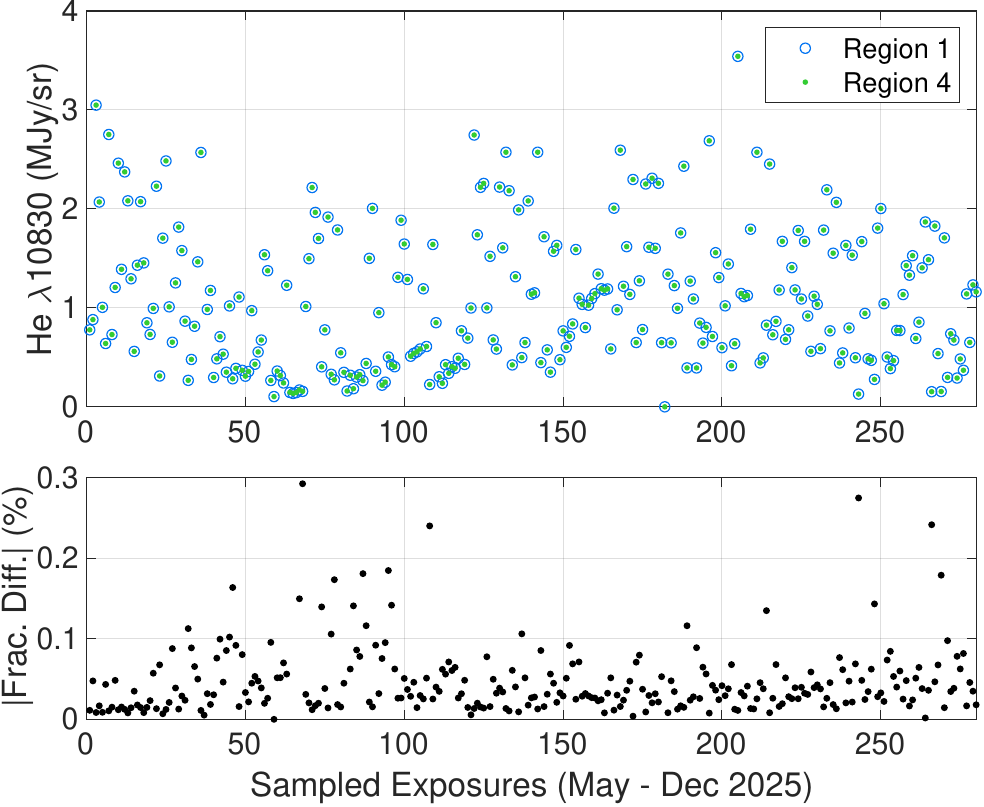}
\caption{
Top: \HeI\ brightness measured independently from the leftmost and rightmost regions in each of the sample exposures.
Bottom: Fractional difference between the \HeI\ brightness measured in these two regions.
The mean difference is $\sim0.05\%$, indicating that the airglow emission is highly spatially uniform on the scale of SPHEREx FOV.
}
\label{fig:example_1D_split_table}
\end{figure}

To evaluate the spatial uniformity, we analyze 280 exposures sampled evenly between May and December 2025.
In each exposure, the detector is divided into four regions along the spatial (horizontal) direction, and a one-dimensional~(1-D) spectrum is extracted from each region.
In Figure~\ref{fig:example_1D_split}, the 1-D spectra from the four regions are compared with the median 1-D spectrum of the full exposure for one representative observation.
For each exposure, \HeI\ is measured independently in each region after background subtraction, following the method described in Section~\ref{sS:multicomponent_fitting}.
The comparison for all 280 exposures are summarized in Figure~\ref{fig:example_1D_split_table}, where the \HeI\ amplitudes in the leftmost and rightmost regions are plotted in the top panel, and their fractional difference is shown in the bottom panel.
The mean fractional difference across the sample is $\sim 0.05\%$, indicating that the \HeI\ airglow brightness is highly uniform within a single exposure.
This result suggests that the percent-level variation seen in Figure~\ref{fig:example_1D_split} is dominated by astronomical background rather than by the spatial variation in the airglow itself.
Furthermore, this uniformity supports the use of a single 1-D spectrum to represent the airglow signal within each exposure.

These results indicate that the line emission varies only weakly across the FOV in the spatial direction.
We therefore integrate over this dimension and treat each exposure as a single integrated measurement of the diffuse airglow, represented by a 1-D spectrum over a $12.25$~deg$^2$ region of sky.
Under this condition, averaging over detector pixels improves signal-to-noise while preserving structure on angular scales larger than the per-exposure FOV.

In this analysis, we use the survey data obtained between 22~April and 31~December of~2025.
During this period, 128,859 calibrated science exposures were obtained.
Observations taken within the South Atlantic Anomaly (SAA) are excluded, as enhanced particle flux produces elevated transient rates that significantly degrade data quality.

%% Masking
\subsection{Masking}
\label{sS:reduction_masking}

Pixel-level quality masks provided with each calibrated exposure are applied to identify non-functioning detector elements and regions contaminated by known bright astronomical sources \cite{Akeson2025}.
While this masking removes the majority of the stellar and Galactic emission, the mask does not fully capture the extended wings of the point spread function~(PSF).
As a result, residual contamination from bright sources can remain in nominally unmasked regions.

An additional sigma-clipping procedure was applied within the wavelength contour bins to mitigate this effect.
Pixels with surface brightness values exceeding the local distribution by more than $3\sigma$ are excluded.
This step suppresses the residual extended PSF contamination while preserving the diffuse background signal.

%% Constructing 1D spectra
\subsection{Constructing one-dimensional Spectra}
\label{sS:reduction_1D}

After masking and outlier rejection, each exposure is collapsed into a 1-D spectrum by binning pixels according to their effective wavelengths.
Within each wavelength bin, the median surface brightness is computed to reduce sensitivity to the small number of remaining contaminated pixels and to provide a robust estimate of the diffuse emission level.
This procedure converts each $2040 \times 2040$ pixel exposure, spanning approximately $3.5 \times 3.5$~deg$^2$, into a single spectrum representing the average sky brightness over that field.

%%%% ---- MODELING AND VALIDATION ---- %%%%
\section{Modeling and Validation}
\label{S:template_fitting}

This section describes the spectral modeling framework used to isolate atmospheric airglow emission from the dominant diffuse background in SPHEREx sky spectra and the criteria used to validate the resulting measurements.
Each exposure is modeled as a linear combination of a Zodiacal light continuum and instrumentally-calibrated emission-line templates, which are fit simultaneously to account for covariance between components.
The construction of the templates, multi-component fitting procedure, estimation of statistical uncertainties, and rejection of exposures with significant non-modeled background contamination are described in the subsections below.

%% Zodi
\subsection{Zodiacal Light Model}
\label{sS:reduction_zodi}

After masking and spectral construction, the remaining signal is dominated by diffuse emission components.
The primary astrophysical foreground in the NIR is Zodiacal light (ZL), which arises from sunlight scattered by interplanetary dust distributed throughout the inner Solar system.

The ZL information used in this analysis is provided as a part of the SPHEREx data products.
These estimates are generated by the SPHEREx sky simulator \cite{Crill2025} using a modified Kelsall model \cite{Kelsall1998}.
The ZL template incorporates the large-scale spatial gradient across the sky, the Solar spectrum, and the ZL spectral variations with ecliptic latitudes.

Rather than subtracting a fixed ZL model, the ZL template is included as a free component in the spectral fit.
For each exposure, the ZL normalization is determined simultaneously with the atmospheric emission line amplitudes.
This approach allows small offsets in the absolute zodiacal light brightness to be absorbed without biasing the inferred airglow signal.

\begin{figure*}
    \centering
    \includegraphics[width=0.95\textwidth]{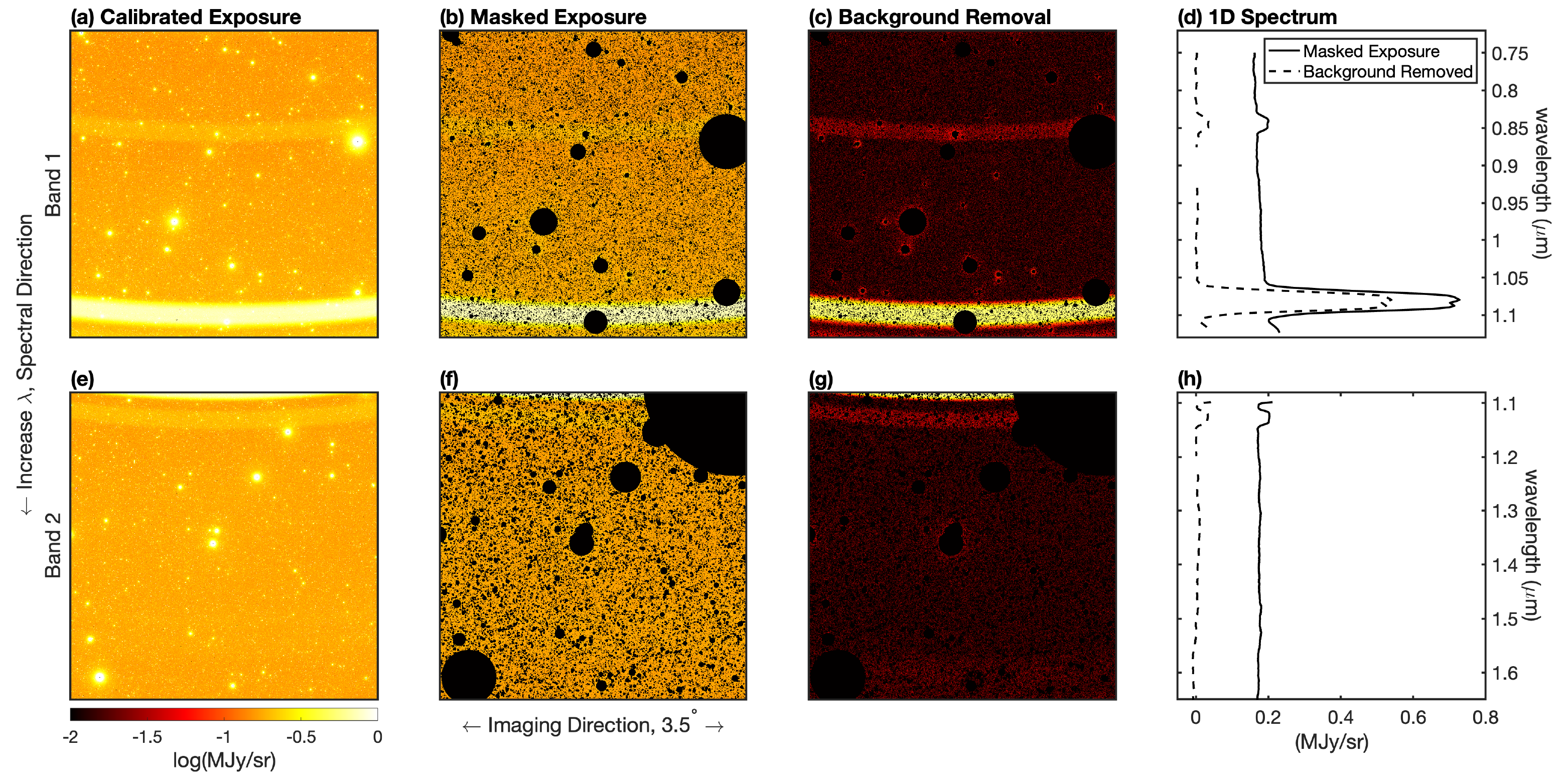}
    \caption{
    Example masked exposures before and after the removal of the best fit Zodiacal light component and the resulting 1-D sky spectra post-removal.
    Top row panels (a to d) are Band~1 examples, while bottom row panels (e to h) are from Band~2.
    (a), (e): input calibrated exposures.
    (b), (f): same exposures with the mask applied.
    (c), (g): the ZL background is removed, showing that the residual signal is dominated by the atmospheric airglow.
    (d), (h): the spectra computed from (c) and (g) by taking the median within each wavelength bin.}
    \label{fig:zodi_subtraction}
\end{figure*}

An example masked exposure is shown in Figure~\ref{fig:zodi_subtraction} before and after subtraction of the best fit ZL component.
After ZL removal, the residual signal is dominated by atmospheric airglow.

%% Line template
\subsection{Atmospheric Emission Line Templates}
\label{sS:lineTemplate}

The spectral templates used in this analysis are derived from the measured SPHEREx spectral response to monochromatic input sources obtained during pre-launch calibration \cite{Hui2024}.

\begin{figure}
    \centering
    \includegraphics[width=0.45\textwidth]{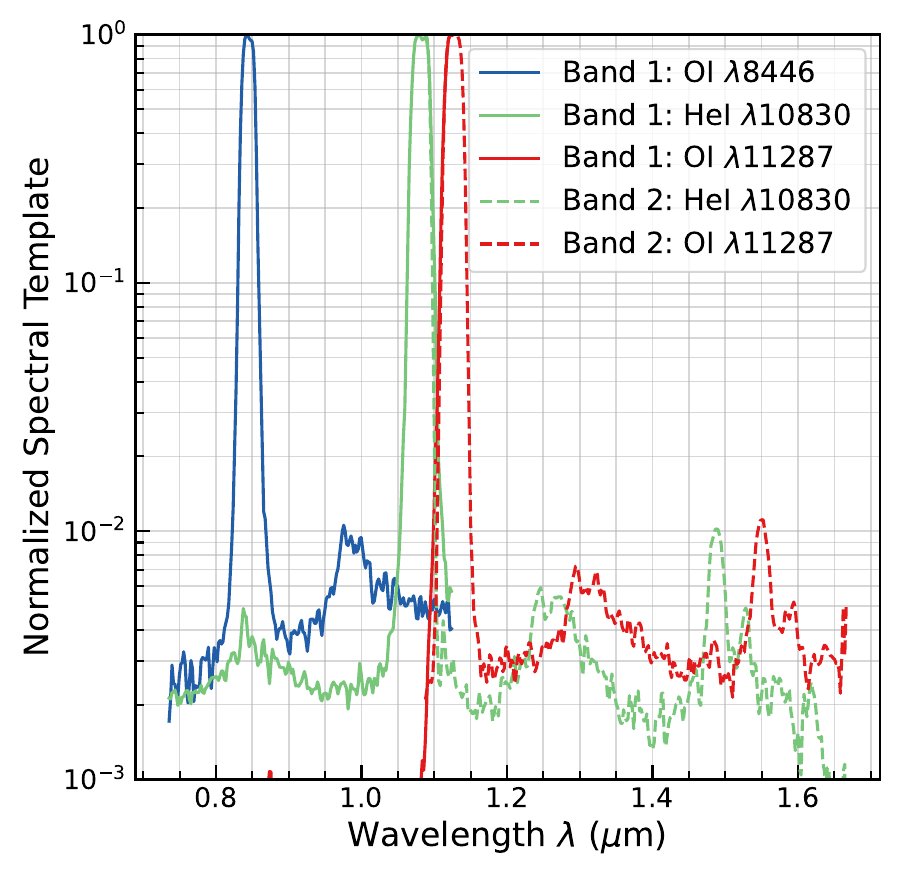}
    \caption{
    The 1-D spectral templates for \OI, \OII, and \HeI\ used in this work.
    The templates are constructed from the measured SPHEREx spectral response to monochromatic input sources.
    From short wavelengths to long wavelength, we show \OI\ (solid, dark blue), \HeI\ in Band~1 (solid, green) and Band~2 (dashed, green), and \OII\ in Band~1 (solid, dark red) and Band~2 (dashed, dark red).
    In Band~1, \HeI\ template includes a leakage feature near 0.845~\micron\ at $\approx$~0.5\% level.
    In Band~2, \HeI\ also has leakage features near 1.244 and 1.486~\micron\ at $\approx$~0.5\% and 1\% levels, respectively.
    The spectral leakage associated with the oxygen templates (green and red) is similarly included but has a negligible effect on the fitted amplitudes.}
    \label{fig:spectral_templates}
\end{figure}

In Figure~\ref{fig:spectral_templates}, we show the 1-D spectral response in Band~1 to monochromatic lines at 0.845, 1.083, and 1.128~\micron, and in Band~2 to lines at 1.083 and 1.128~\micron.
These responses define the templates for the \OI, \OII, and \HeI\ airglow features analyzed in this work.

SPHEREx filters in this wavelength range exhibit known spectral leakage \cite{Hui2026}.
The associated leakage features have a spectral bandwidth and shape that differ from the nominal SPHEREx spectral response function to a monochromatic input and are narrower than that expected for the native resolving power of $R \sim 41$.
At 1.083~\micron, \HeI\ emission produces a leakage feature at $\sim 0.5\%$ near 0.838~\micron\ in Band~1 and $\sim 1\%$ near 1.479~\micron\ in Band~2.

Although the leakage has a distinct spectral profile, the feature at 0.838~\micron\ lies close to the \OI\ line.
Because the helium airglow is typically the dominant diffuse emission component in low Earth orbit and can exceed the oxygen emission by up to two orders of magnitude, this leakage feature is explicitly included in the helium template to avoid biasing the oxygen line measurements.
To verify that the leakage component does not bias the recovered oxygen signal, we performed injection tests in which the amplitude of the \HeI\ emission was varied and the modified spectra were processed through the same fitting pipeline.
The recovered \OI\ emission amplitude remains consistent with the input value across these tests.

The spectral templates for \OI\ and \OII\ similarly include their associated leakage features.
The oxygen emission is significantly weaker, and leakage from these oxygen lines is well below the detection threshold in this analysis.
Nevertheless, these features are included for completeness and consistency.

%% Multi-component fitting
\subsection{Multi-component Spectral Fitting}
\label{sS:multicomponent_fitting}

Each 1-D spectrum is modeled as a linear combination of the ZL continuum and the atmospheric emission line templates.
All components are fit simultaneously with free amplitudes.

\begin{figure*}
    \centering
    \includegraphics[width=0.98\textwidth]{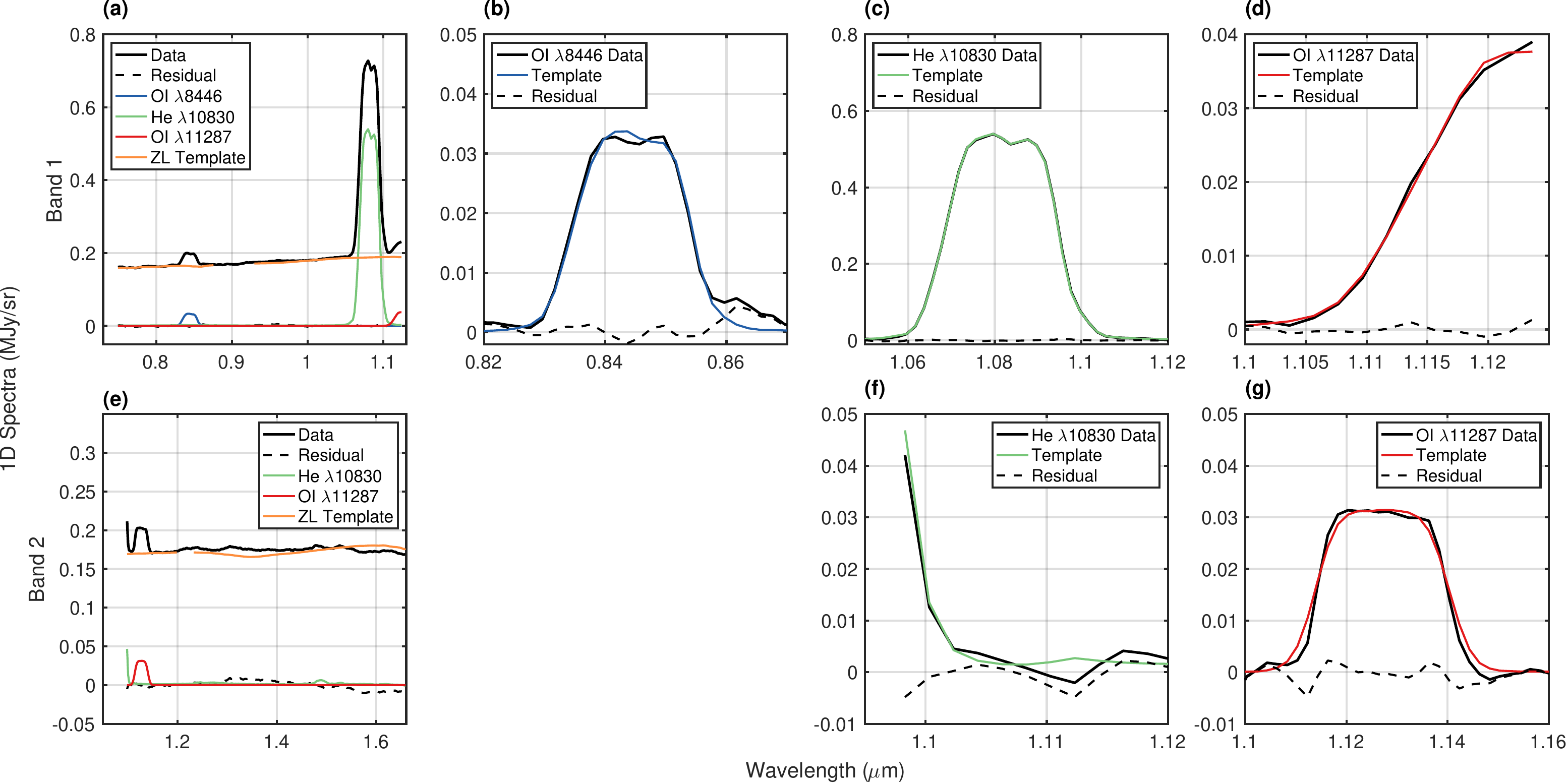}
    \caption{
    Example multi-component template fitting results for a single SPHEREx exposure.
    The top row panels (a to d) represent result from Band-1, while the bottom row panels (e to g) are results from Band-2.
    \textit{Top}: (a) Band~1 observed 1-D spectrum shown with the best fit combination of the ZL continuum and the atmospheric emission line templates.
    (b) Band~1 best fit template for the O\textsc{i} 0.845~\micron\ line.
    The data after subtraction of all other fitted components are shown, with the dotted line indicating the residual.
    (c) and (d) Same as in (b), but for the He\textsc{i} 1.083~\micron\ and O\textsc{i} 1.128~\micron\ lines in Band~1.
    \textit{Bottom}: (e) Band-2 observed 1-D spectrum with the best fit combination, similar to (a).
    (f) and (g) are the same lines as in (c) and (d).
    }
    \label{fig:band_fitting}
\end{figure*}

For Band~1, the model spectrum is expressed as:
\begin{eqnarray}
I_{\mathrm{Band1}}(\lambda) &=&
\alpha_{\mathrm{Zodi}}\,T_{\mathrm{Zodi}}(\lambda) \nonumber \\
&+& \alpha_{\mathrm{OI845}}\,T_{\mathrm{OI845}}(\lambda) \nonumber \\
&+& \alpha_{\mathrm{He1083}}\,T_{\mathrm{He1083}}(\lambda) \nonumber \\
&+& \alpha_{\mathrm{OI1128}}\,T_{\mathrm{OI1128}}(\lambda) \nonumber \\
&+& r(\lambda),
\end{eqnarray}
where $T(\lambda)$ denotes the normalized templates, $\alpha$ are the fitted amplitudes, and $r(\lambda)$ are the residuals.
In Figure~\ref{fig:band_fitting}, we include an example Band~1 exposure and the corresponding multi-component template fit.

The amplitudes are determined using a linear least squares fit.
Simultaneous fitting of the ZL and emission line templates is required.
Otherwise, small mismatches in ZL normalization from bright emission lines would bias the inferred airglow amplitudes if the continuum were removed \textit{a priori}.
In addition, spectral leakage from the bright emission lines, particularly \HeI, introduces coupling between the line templates and the continuum level.
Fitting all components together allows these covariances to be absorbed in a self-consistent manner by the model.

The same fitting framework is applied to Band~2 using the corresponding templates.
Because the instrumental line spread function extends across the Band~1 and Band~2 boundary, \HeI\ and \OII\ responses are partially sampled in both bands.
These partial responses are included in the templates and fit simultaneously to maintain internal consistency.

\begin{figure}
    \centering
    \includegraphics[width=0.46\textwidth]{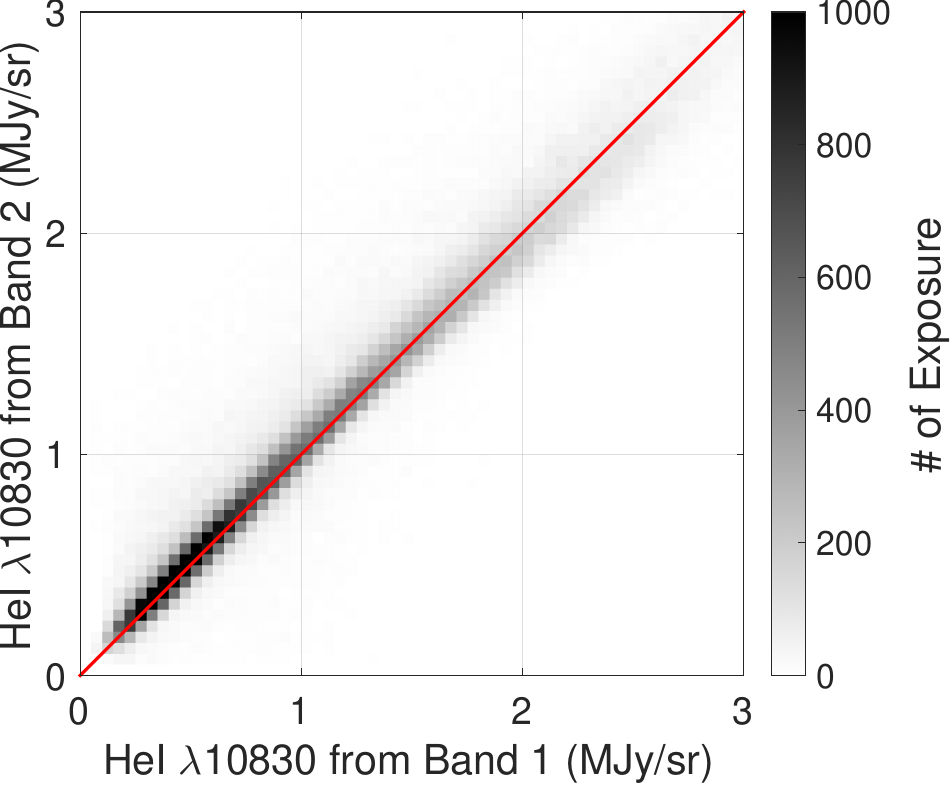}
    \caption{Comparison of the fitted emission line amplitudes derived independently for Band~1 and Band~2 for \HeI.
    The recovered amplitudes are broadly consistent between the two bands.
    In Band~2, the lowest spectral bin begins at 1.09\micron\ (Figure~\ref{fig:band_fitting}), so the fit constrains only the tail of the \HeI\ line response.
    As a result, the Band~2 amplitudes show larger scatter than those from Band~1.
    \label{fig:band1_band2_consistency}}
\end{figure}

For quantitative measurements and assessments of the signal significance, Band~1 result is adopted for \HeI\ and Band~2 result is adopted for \OII, where each feature is best constrained by the respective band spectral response.
The inclusion of complementary band information primarily improves robustness against template mismatch and continuum fitting uncertainties.
To show the measurement consistency between the two bands, the independently fitted \HeI\ amplitudes derived from Band~1 and Band~2 are compared in Figure~\ref{fig:band1_band2_consistency}.

Most exposures in which oxygen emission is detected show measurable signal in both the \OI\ and \OII\ lines.
However, a small fraction of exposures show a detection in only one band, either \OI\ in Band~1 without a corresponding \OII\ detection in Band~2, or vice versa.
This mismatch could arise from differences in masking and spatial coverage between the bands, differences in signal-to-noise and spectral response, or from physical differences in the excitation mechanisms that produce the two.

\begin{figure}
    \centering    
    \includegraphics[width=0.46\textwidth]{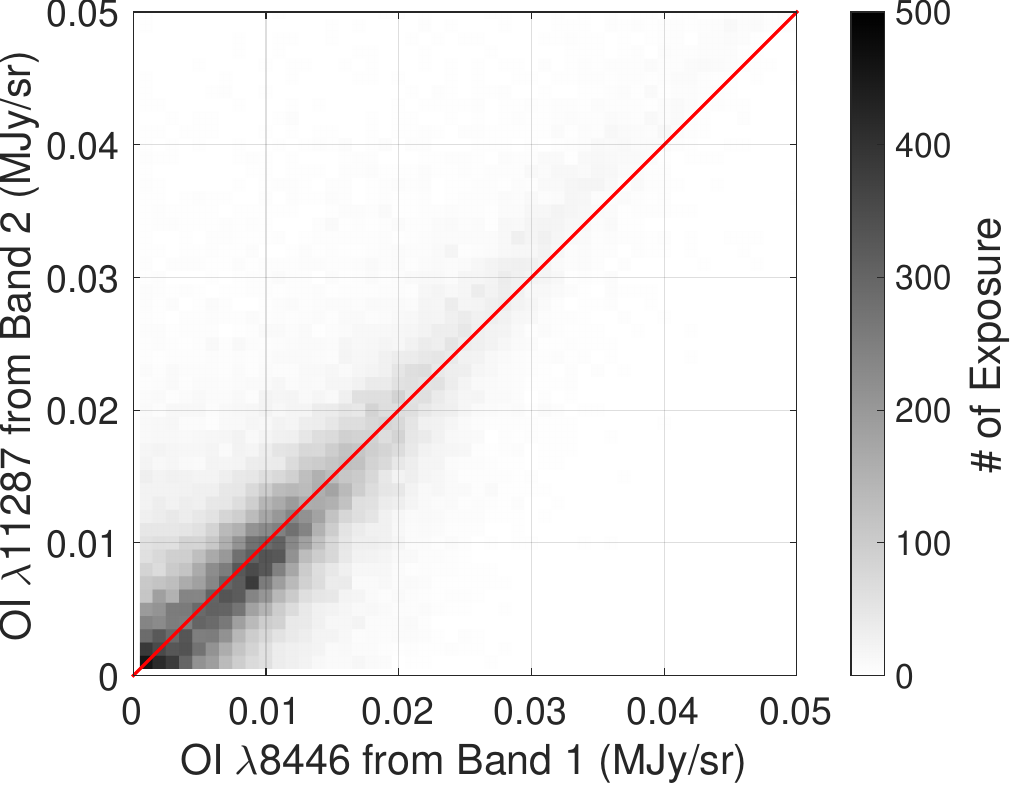}
    \caption{Comparison of \OI\ brightness from Band~1 and \OII\ brightness for Band~2.
    Most exposures follow a tight correlation consistent with a common excitation pathway.
    The solid, red line indicates a relationship with a slope of one and passes through the origin. 
    We fit the actual relationship between \OI\ and \OII\ and find a slope of 0.973 for a line passing through the origin.
    %The brightest \OII\ measurements that deviate from this trend are associated with intervals of elevated geomagnetic activity discussed in Section~\ref{sS:solar_activity}.
    }
    \label{fig:oxygen_consistency}
\end{figure}

To examine the relation between the two oxygen lines when both are detected, we compare the fitted amplitudes of \OI\ and \OII\ for exposures where both lines are measured, as shown in Figure~\ref{fig:oxygen_consistency}.
The two oxygen features are generally detected consistently and show a strong correlation.

We evaluate the relation between \OI\ and \OII\ amplitudes during quiescent solar activity periods using linear regression with Huber loss function to reduce the influence of outliers.
For a fit constrained to pass through the origin, we find an \OII\ versus \OI\ slope of~0.973~$\pm$~0.001.
Allowing both slope and intercept gives a slope of 0.939~$\pm$~0.002 and a positive but small intercept (0.00048 $\pm 2\times10^{-5}$), indicating a modest departure from a purely proportional relation between the two lines near zero signal.
In both cases, the \OI\ amplitude is on average larger than the \OII\ amplitude.

For the analysis presented in Section~\ref{S:discussion}, we use only exposures in which both Band~1 and Band~2 oxygen lines are detected, so that the comparison between the two lines is not affected by single band non-detections.

%% Errorbar
\subsection{Detection Uncertainty}
\label{sS:expUncert}

The uncertainties on the fitted emission line amplitudes are dominated by the modeling uncertainties rather than the detector noise.
Each wavelength bin averages over several thousand pixels, rendering the instrumental noise subdominant.
Instead, the main contributions comprise imperfect modeling of the diffuse background, primarily ZL, and the residual astrophysical contamination in the exposures.

The detection uncertainties are estimated from the residuals in a local wavelength window around each emission feature, after subtraction of the best fit model.
For a line centered at wavelength $\lambda_0$, the bandwidth is defined by the resolving power $R$:
\begin{equation}
\Delta\lambda = \frac{\lambda_0}{R}.
\end{equation}
For \OI\ in Band~1 with $R \approx 41$, this corresponds to $\Delta\lambda \approx 0.021~\mu\mathrm{m}$.

The residuals are evaluated in a wavelength window spanning $\pm 1.5\,\Delta\lambda$ around the line center.
For \OI, this corresponds to:
\begin{equation}
\lambda \in [0.824,0.866]~\mu\mathrm{m}.
\label{Eq:lambda_window}
\end{equation}
Within this window, the residual spectrum is defined as:
\begin{equation}
r(\lambda) = I(\lambda) - \sum_j \alpha_j T_j(\lambda),
\end{equation}
where $I(\lambda)$ is the observed 1-D spectrum, $T_j(\lambda)$ are the normalized spectral templates described in Section~\ref{sS:lineTemplate}, and $\alpha_j$ are the best fit amplitudes obtained from the multi-component fitting procedure in Section~\ref{sS:multicomponent_fitting}.

Given this residual spectrum, its variance is estimated as:
\begin{equation}
\hat{\sigma}^2 =
\frac{1}{N-1}
\sum_{\lambda} r^2(\lambda),
\end{equation}
where the sum runs over the $N$ wavelength bins within the window defined in Equation \ref{Eq:lambda_window}.

For a single template amplitude $\alpha_k$ measured in this window, the corresponding uncertainty is estimated as:
\begin{equation}
\sigma_{\alpha_k} =
\sqrt{
\frac{\hat{\sigma}^2}
{\sum_{\lambda} T_k^2(\lambda)}
},
\end{equation}
where $T_k(\lambda)$ is the template for the emission line of interest sampled on the same wavelength grid.

An emission line is considered detected when the fitted amplitude satisfies $\alpha_k / \sigma_{\alpha_k} \ge 3$.
In practice, detections of \HeI\ are based on the Band~1 fitted amplitude, where the instrumental response provides the strongest constraint.
For the O\textsc{i} emission, a detection is required in both \OI\ measured in Band~1 and \OII\ measured in Band~2, with each satisfying $\mathrm{SNR} \ge 3$ in the same observation.

The quoted uncertainties are empirical estimates derived from the local residual variance and assume negligible covariance between templates.
They represent statistical detection uncertainties and do not constitute a full propagation of the multi-component fit covariance.

Several sources of systematic uncertainty are not included in the empirical detection uncertainties reported above.
These include uncertainties in the absolute flux calibration, small variations in the spectral response function, and uncertainties in the calibration of the spectral leakage features.
These effects primarily impact the absolute brightness scale rather than the statistical significance of the line detection.
Based on the calibration analyses presented in \citeA{Korngut2026, Hui2026}, we estimate that these systematic effects are at the level of approximately $5\%$ in the recovered line amplitudes.

%% Background contamination
\subsection{Background Contamination Rejection}
\label{sS:fitValidation}

A subset of SPHEREx exposures contains strong astrophysical background emission that is not adequately described by the ZL model.
This emission is dominated by regions of high stellar density and the diffuse Galactic light, primarily toward the Galactic bulge, the Galactic plane, and the Large Magellanic Cloud (LMC).
In order to simplify the spectral modeling framework adopted in this analysis, such exposures are excluded.
In addition, because auroral emission templates are not included in the present modeling framework, exposures dominated by strong broadband auroral emission are also removed.

These cases are identified using a goodness of fit (GOF) metric that quantifies how closely the observed background spectrum follows the expected ZL spectral shape.
For each exposure, the ZL model is first fit in amplitude to the observed background spectrum.
The GOF is then evaluated using the normalized~$\chi^2$,
\begin{equation}
\chi^2 = \frac{1}{N}
\sum_{i}
\frac{\left[ I(\lambda_i) - \alpha_{\mathrm{ZL}}T_{\mathrm{ZL}}(\lambda_i) \right]^2}
{\sigma^2(\lambda_i)},
\end{equation}
where $\sigma(\lambda_i)$ is the per channel uncertainty.
The sum is performed over wavelength regions dominated by diffuse background emission and excludes channels affected by strong atmospheric emission lines.

When the background is well described by the ZL model, the normalized chi squared values cluster near unity.
Exposures that deviate significantly from the ZL spectral shape exhibit elevated $\chi^2$ values and are rejected when $\chi^2 > 2$, indicating poor agreement with the ZL continuum.

\begin{figure}
    \centering
    \includegraphics[width=0.48\textwidth]{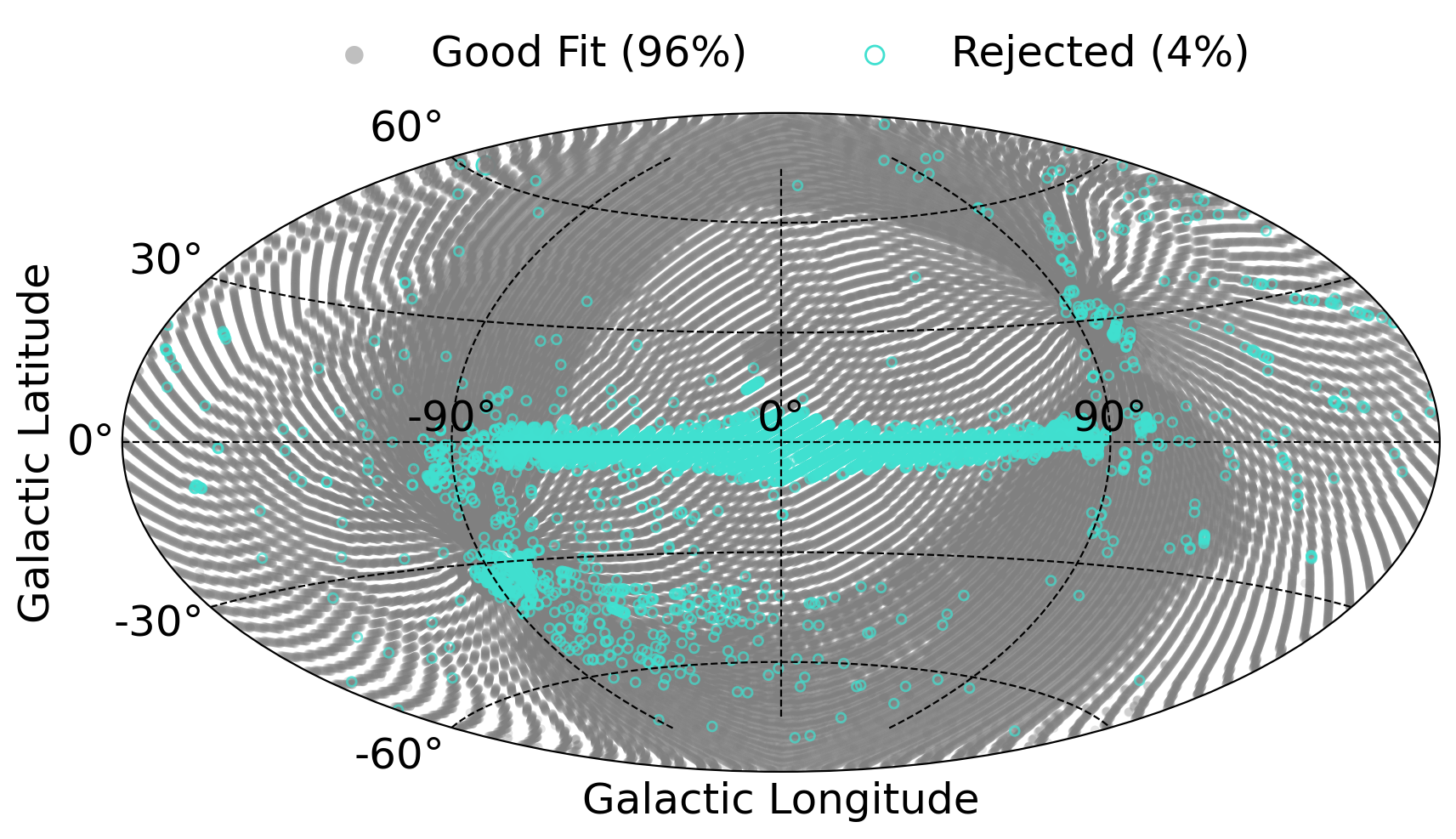}
    \caption{
    The sky location of the accepted and rejected fit results after the goodness of fit cut in Galactic coordinates of SPHEREx pointing, highlighting that the rejected exposures are concentrated along the Galactic plane and near the LMC.
    In these regions, strong astrophysical backgrounds would otherwise bias the fitted atmospheric emission line amplitudes.
    The numbers in parentheses denote the fractions of the total data set.}
    \label{fig:gof_sky_mask}
\end{figure}

This cut removes approximately $10\%$ of the calibrated exposures and is strongly clustered in Galactic coordinates.
In Figure~\ref{fig:gof_sky_mask}, the two-dimensional sky distribution of accepted and rejected exposures is presented.

%%%% ---- TABULATED RESULT ---- %%%%
\section{Airglow Brightness Results}
\label{S:result}

\begin{figure*}
	\centering
    \includegraphics[width=0.32\textwidth]{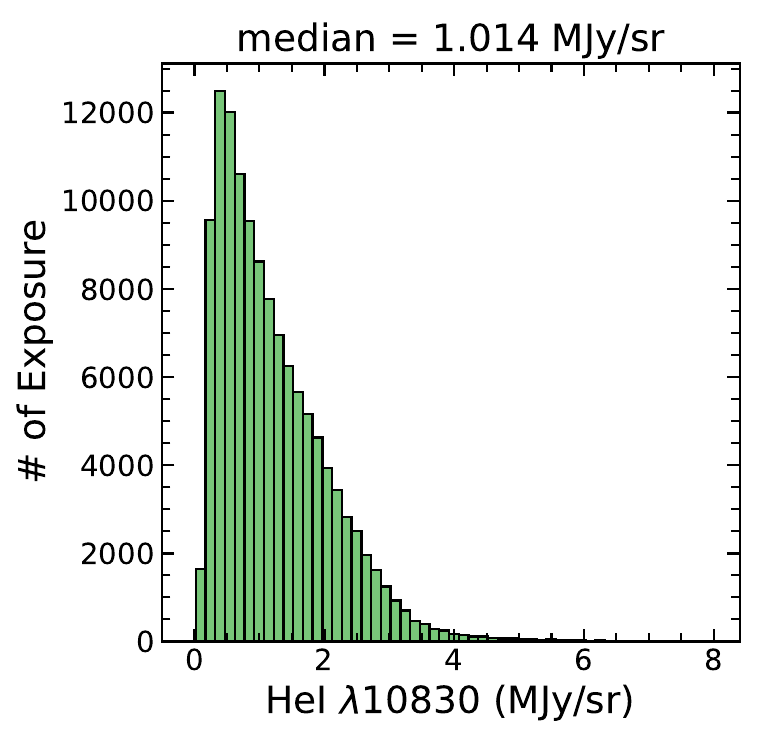}
    \hfill
    \includegraphics[width=0.32\textwidth]{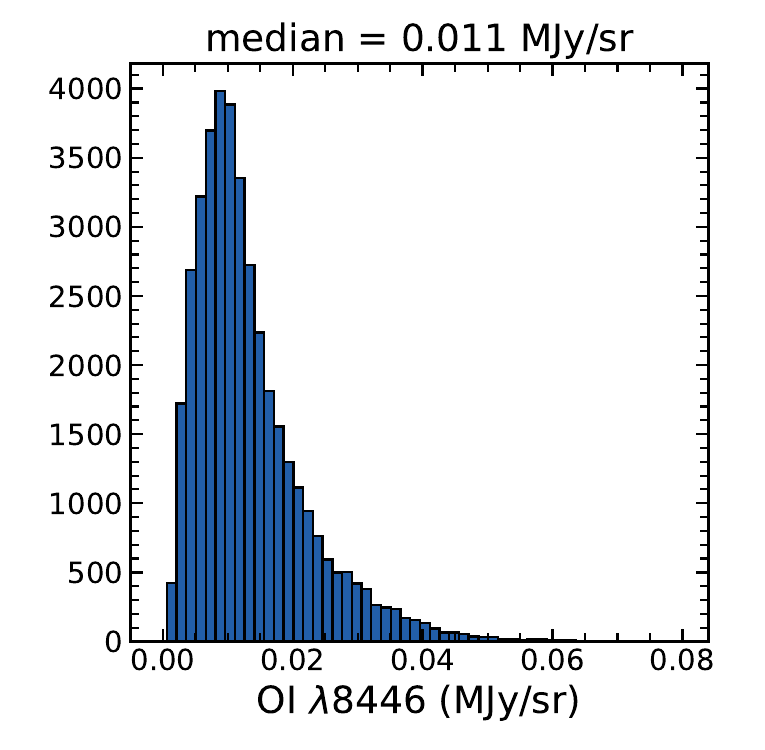}
    \hfill
    \includegraphics[width=0.32\textwidth]{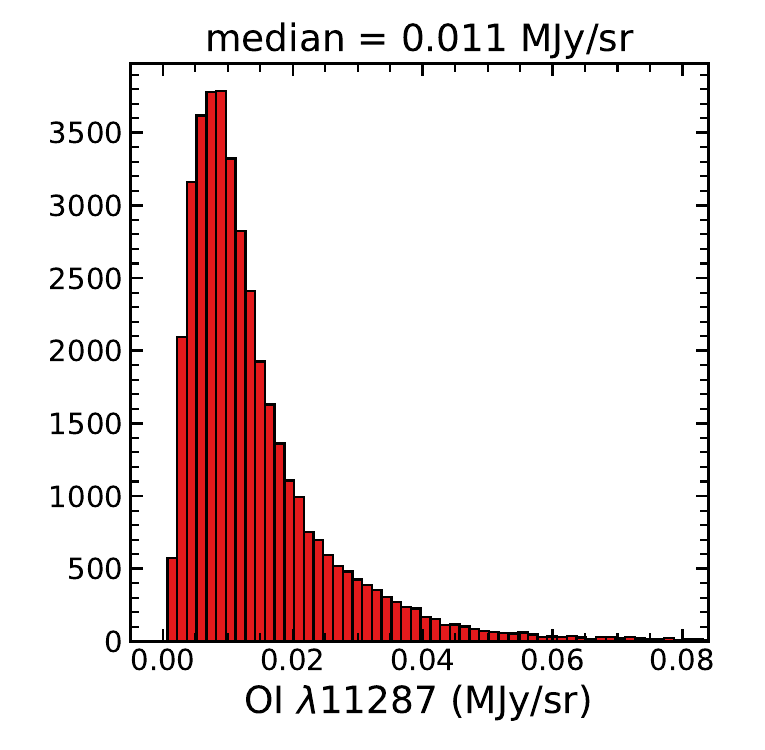}
	\caption{
	Distributions of fitted airglow amplitudes.
    (a) The histogram of 122,608 \HeI\ detections binned by 0.15~MJy/sr.
    (b) The histogram of 39,500 \OI\ and (c) 39,500 \OII\ detections at 0.0015~MJy/sr binning.
    The median helium amplitude 1.014~$\pm$~0.001~MJy/sr is approximately 10$\times$ the typical ZL amplitude of about $10^{-1}$~MJy/sr. 
    The median oxygen amplitudes are 1.11~$\pm$~0.06$\times$10$^{-2}$ and 1.08~$\pm$~0.06$\times$10$^{-2}$~MJy/sr for \OI\ and \OII, respectively.
    The quoted uncertainties represent the error on the median and do not reflect the dispersion of the distribution.
\label{fig:DataStat}}
\end{figure*}

After applying the detection and rejection criteria, the remaining data set provides near-continuous temporal coverage of helium and oxygen airglow over an eight-month survey period.
% In Table \ref{tab:med_brightness}, we summarize the median brightness and uncertainties of all three emission lines.
The histograms of the validated fitted amplitudes for the helium and oxygen airglow are presented in Figure \ref{fig:DataStat}.
Most notably, the median brightness of \HeI\ exceeds the median ZL background at this wavelength by roughly an order of magnitude.
The large number of accepted exposures enables characterization of variability on both orbital and seasonal timescales, which is discussed in Section \ref{S:discussion}.

%%%% ---- DISCUSSION ---- %%%%
\section{Discussion}
\label{S:discussion}

\begin{figure*}
    \centering
    \includegraphics[width=0.95\textwidth]{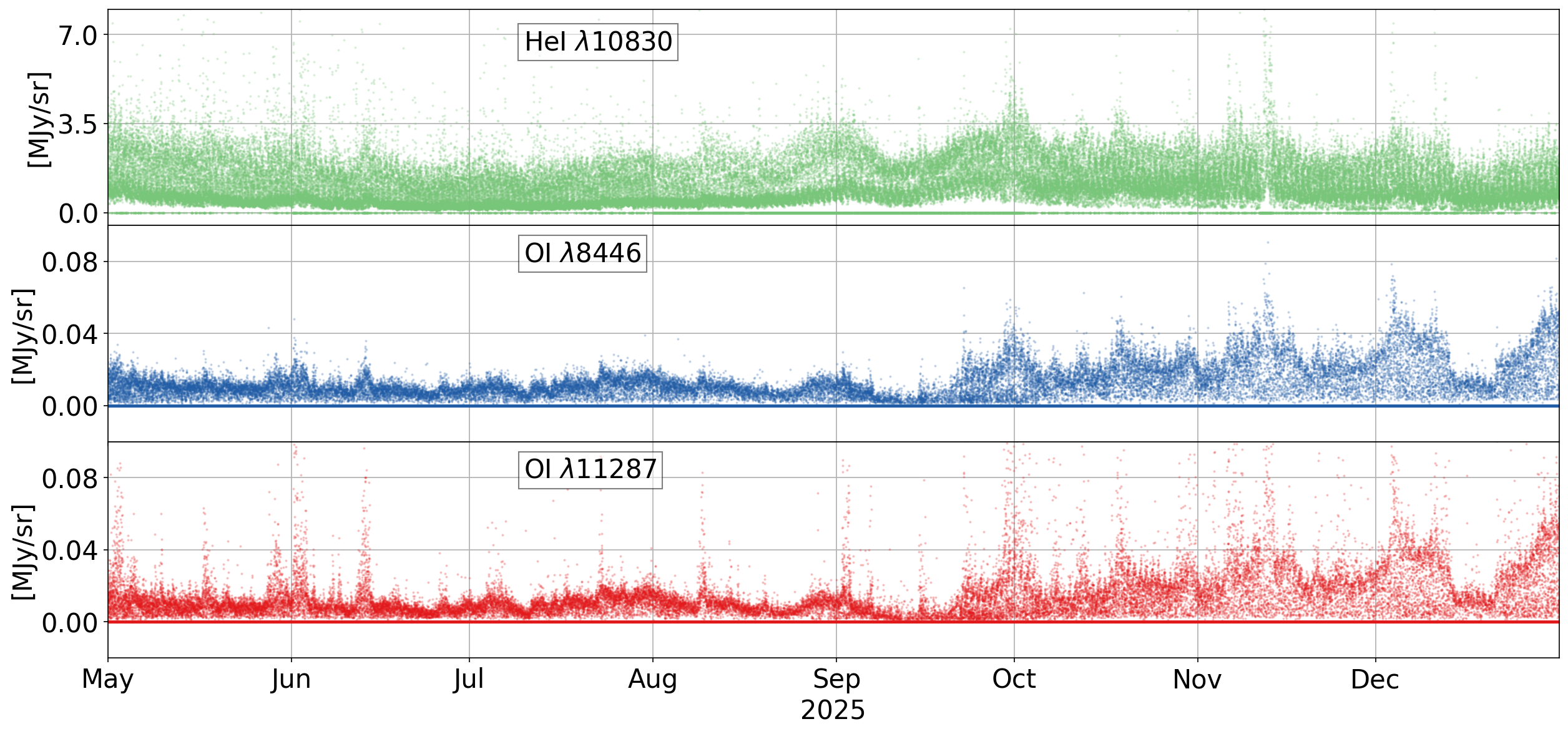}
    \caption{
    The time series of the fitted atmospheric airglow amplitudes from May through December 2025 where every data point is a detection.
    From top to bottom panels: \HeI, \OI, and \OII.
    The amplitudes exhibit both short- and long-timescale temporal variations.
    In addition, other factors like SPHEREx location, Solar illuminating geometry, and Solar activity can affect the measured line brightness. 
    }
    \label{fig:fullData}
\end{figure*}

In this section, we summarize the dominant trends in the fitted airglow amplitudes and relate them to geographic location, SZA, and geomagnetic activity.
In Figure~\ref{fig:fullData}, the airglow brightness is plotted as a function of time from May to December 2025 where each point represents a single exposure.
Variability can be seen on multiple timescales, including orbit-scale modulation, seasonal evolution, and episodic enhancements coincident with elevated geomagnetic activity.

To account for the geomagnetic conditions, the planetary Kp index (https://kp.gfz.de/)~\cite{Matzka2021} at the time of each SPHEREx observation is adopted as a coarse tracer.
The Kp index is a three hour global geomagnetic activity index derived from ground based magnetometer measurements and provides a measure of disturbances in the Earth magnetic field driven by solar wind variability.
A quiet sample is defined to have $\mathrm{Kp} < 3$ and is used to characterize the baseline atmospheric structure.
Observations during periods of higher Kp values are retained only for the purpose of illustrating storm-time enhancements.

%% Solar activities
\subsection{Enhancement from Solar Activity}
\label{sS:solar_activity}

Enhancement from Solar activity is accompanied by additional energy deposition and ionization in the upper atmosphere, which can modify airglow emission.
The spectral model does not include an explicit component for auroral emissions, which are observed in SPHEREx data as multiple spectral lines that become blended together at the filter resolution.
As a result, the exposures dominated by aurora typically fail the fit validation described in Section~\ref{sS:fitValidation} and are excluded from the validated sample.
However, moderate enhancements of the line features due to temporary changes in the local atmospheric conditions following a geomagnetic storm can still be captured in SPHEREx data.

\begin{figure*}[htb]
	\centering   
    \includegraphics[width=0.32\linewidth]{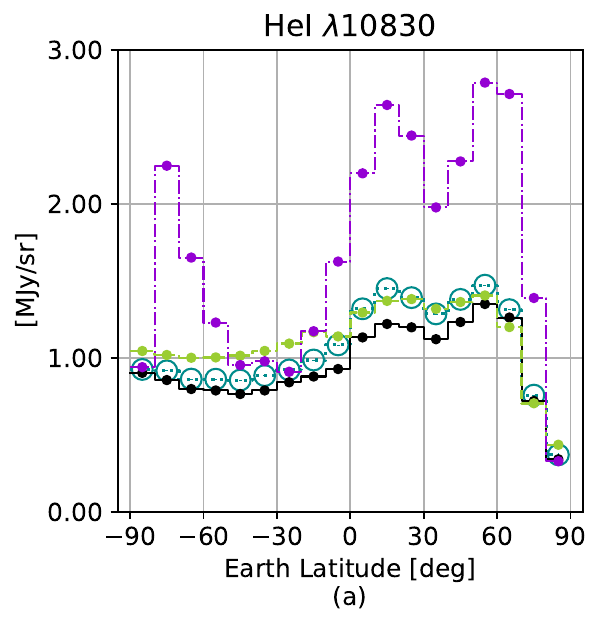}
    \hfill
    \includegraphics[width=0.32\linewidth]{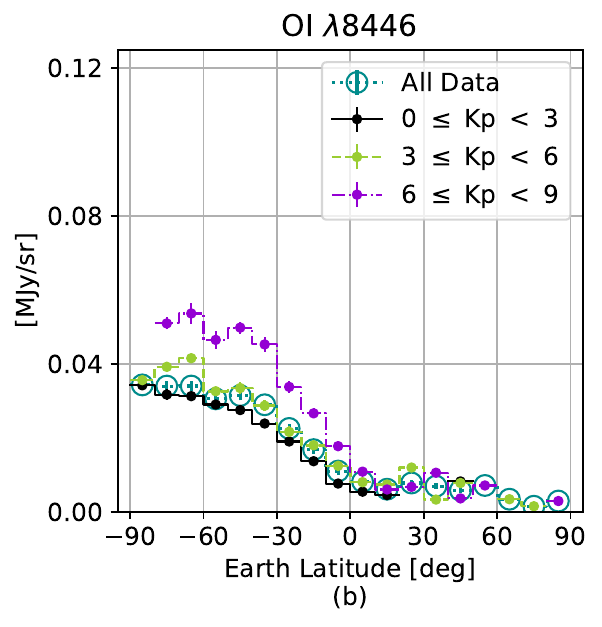}
    \hfill
    \includegraphics[width=0.32\linewidth]{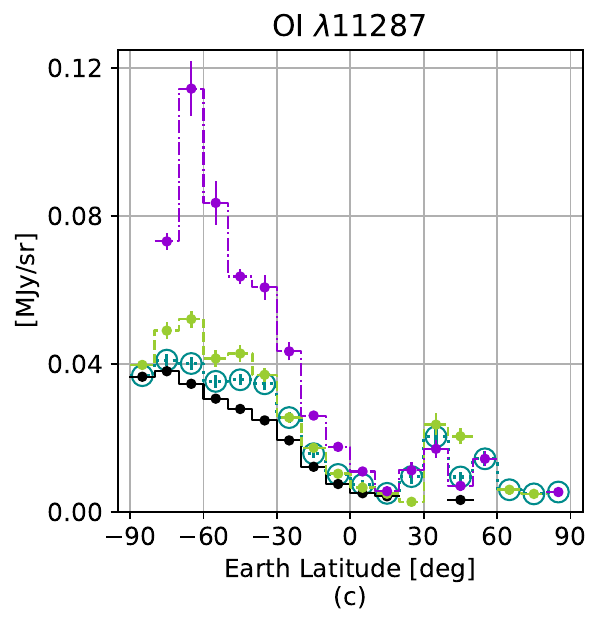}
	\caption{   
    The average brightness of the helium and oxygen airglow as a function of terrestrial latitudes between 9-16~November~2025, covering the time window before and after a strong Solar storm on 12~November~2025.
    (a) \HeI, (b) \OI,(c) \OII.
    To examine the influence of geomagnetic activity, the data are separated by the planetary Kp index as a proxy for global geomagnetic disturbance: 0 $\leq$ Kp $<$ 3 (solid black line), 3 $\leq$ Kp $<$ 6 (dashed light green), 6 $\leq$ Kp $<$ 9 (dash dotted purple).
    Enhanced emission is observed in the data taken during elevated geomagnetic activities (Kp $>$ 3), raising the global average of the brightness (dotted dark green line).
	\label{fig:Kp_trend}}
\end{figure*}

\begin{figure*}[htb]
	\centering   
    \includegraphics[width=0.32\linewidth]{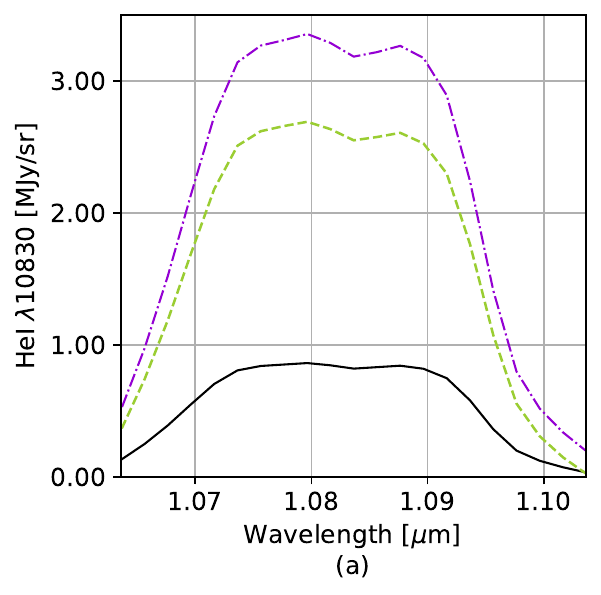}
    \hfill
    \includegraphics[width=0.32\linewidth]{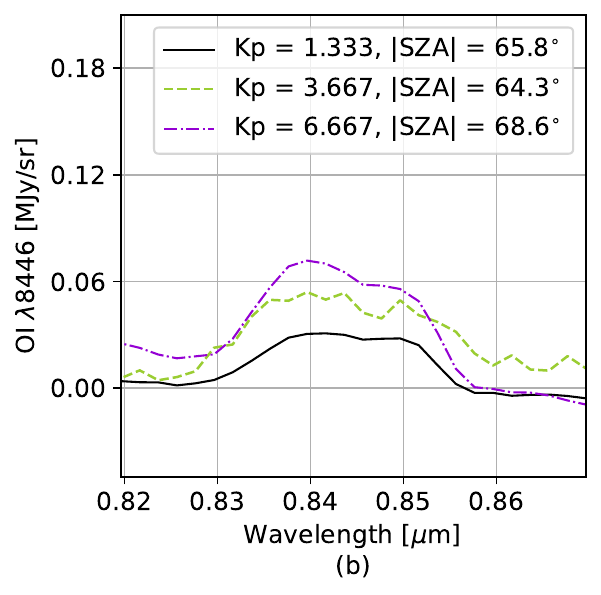}
    \hfill
    \includegraphics[width=0.32\linewidth]{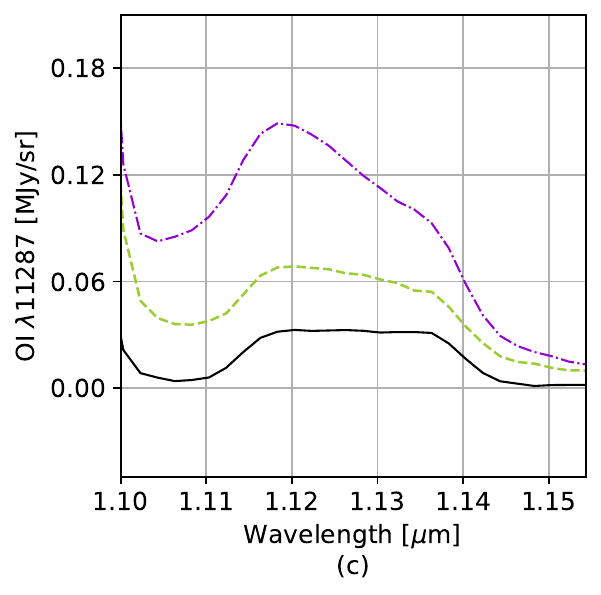}
	\caption{
    Example 1-D spectra from three exposures taken on 13~November~(Kp~$>$~3) and 15~November~(Kp~$<$~3).
    The spectra are zoomed into the wavelength windows that are used in the model: (a) \HeI, (b) \OI, and (c) \OII.
    The three exposures were taken over the same region of the atmosphere, with terrestrial latitude range $-73^{\circ}$ to $-61^{\circ}$ and longitude range $-129^{\circ}$ to $-125^{\circ}$. 
    The best-fit ZL model is subtracted from each spectrum to remove variations from the astrophysical background.
    Given the small latitude/longitude range and the short time separation between the observations, the large variations in the line brightness are likely associated with the Solar storm that arrived on 12~November.
	\label{fig:Kp_1D_spectra}}
\end{figure*}

During the geomagnetically disturbed interval in November 2025 that reached Kp=9, all three emission lines exhibit increased brightness relative to quiet conditions, although the enhancement differs between the lines and the latitude distribution as shown in~Figure~\ref{fig:Kp_trend}.
An example of the line enhancement is found in Figure~\ref{fig:Kp_1D_spectra}.
The ZL-subtracted 1D spectra are constructed from three exposures taken over the same region of the atmosphere on 13~November (Kp~$>3$) and 15~November (Kp~$<3$) under different geomagnetic activity levels.
After removing the dominant sky background by subtracting the ZL continuum, the variations in the line emission observed in this region over a short timescale are most likely associated with geomagnetic disturbances caused by the Solar storm on 12~November.

In the case of helium, the flux at the poles and between $0 - 30^{\circ}$ increases with higher Kp indices.
For oxygen, appreciable increase in brightness is only observed at the South pole $\leq30^{\circ}$ and no similar increase is found in the aurora circle at the North pole.
The brighter helium emission at the poles may suggest influences of aurora, because the enhancement is observed within the known North and South aurora circles.
On the other hand, the excess oxygen emission near the South pole and the excess helium emission near the equator is not likely to be aurora, but still has some dependencies on geomagnetic activity.
A possible explanation is that themospheric heating from the storm increases the gas density at SPHEREx altitude temporarily, leading to a boost in the detected line amplitude \cite{Wang2025, Delano2024, Zesta2019}.

Similar short-lived enhancements are present at several epochs throughout the eight-month survey (Figure~\ref{fig:fullData}), coincident with elevated geomagnetic activity.
Under quiet conditions (Kp~$<3$), \OI\ and \OII\ remain tightly correlated across latitude and time.
However, during disturbed intervals, \OII\ exhibits a systematically larger enhancement than \OI\ within the same exposures.
Figure~\ref{fig:Kp_trend}c shows that the spectral region between 1.10 and 1.11~\micron\ is enhanced during periods with $\mathrm{Kp}>6$, indicating that additional unresolved emission features may contribute within or near the observed \OII\ bandpass at the SPHEREx spectral resolution.
Therefore, we cannot infer from these measurements alone that \OII\ is intrinsically larger relative to \OI\ during disturbed conditions.

To examine the underlying atmospheric structure in the absence of geomagnetic perturbations, all subsequent analysis is restricted to observations obtained under quiet conditions $\mathrm{Kp} < 3$.

%% Orbital trends
\subsection{Orbital Variability and Geometric Coupling}
\label{sS:orbital_trend}

\begin{figure}
    \centering
    \includegraphics[width=0.48\textwidth]{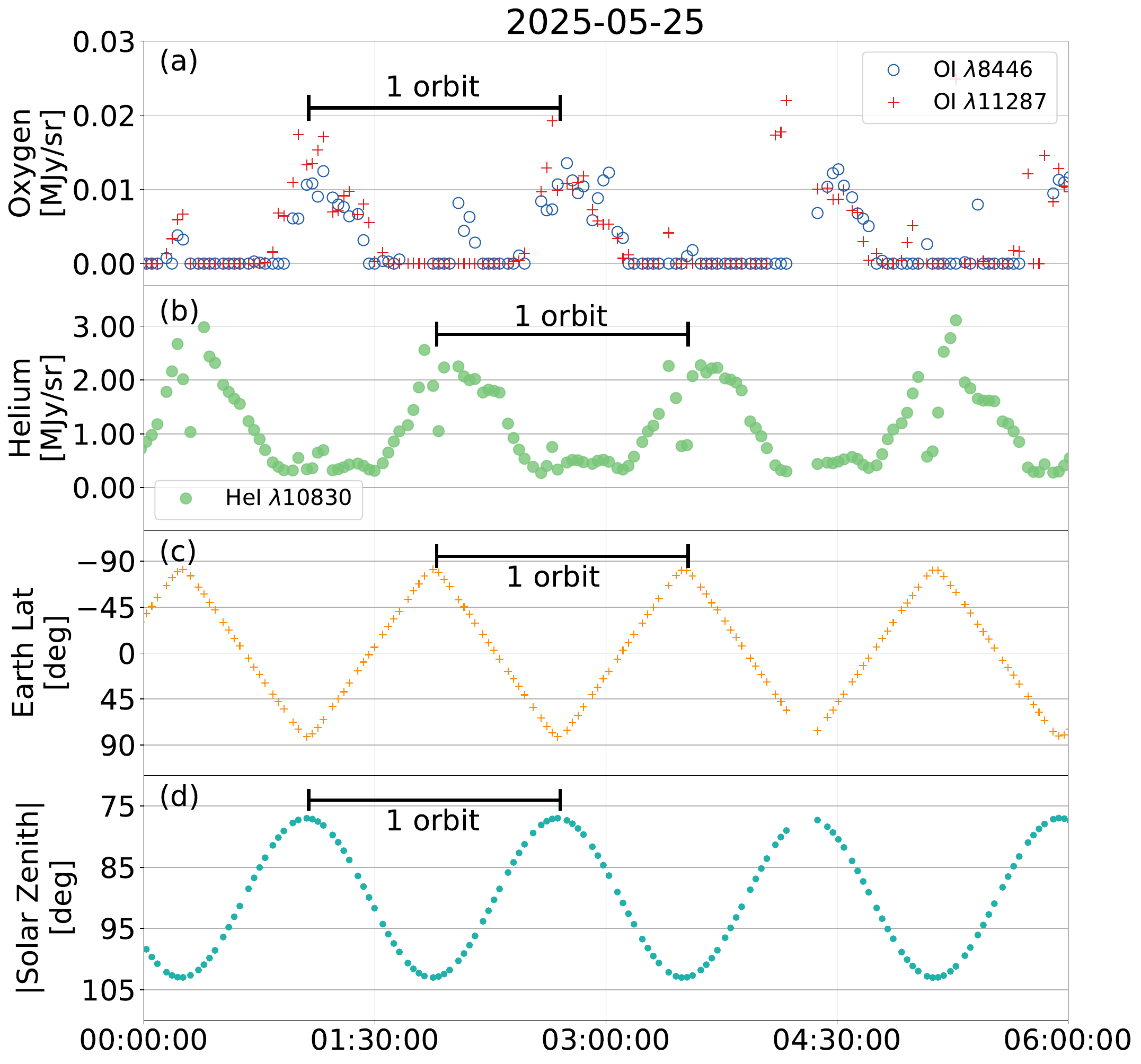}
    \caption{Six-hour time series of the airglow brightness on 25 May 2025. 
    (a) \OI\ and \OII\ amplitudes. 
    (b) \HeI\ amplitude. 
    (c) The ground-tracking latitude of the spacecraft during an observation.
    (d) The absolute value of the SZA, representing how close the pointing was to the Sun.
    The sign of the SZA denotes which half of the orbit SPHEREx is on (sunrise/sunset) and is not considered in this work.
    All three lines exhibit variability on a $\sim$98-minute cadence, matching the SPHEREx orbital period.
    While at a glance the helium brightness (panel a) appears to anti-correlate with oxygen (panel b), this apparent trend is not evident of a direct relationship between the two gas species.
    The helium brightness varies more closely with spacecraft latitude, while the oxygen brightness tracks the SZA.
    In panels (c) and (d), the coupling between latitude and absolute value of the SZA within an orbit is evident, which limits their independent interpretation and contributes to the apparent anti-correlation of helium and oxygen.    \label{fig:orbital}
    }
\end{figure}

Beyond geomagnetic enhancements, the airglow amplitudes exhibit systematic variability on orbital timescales.
A modulation with a cadence matching the SPHEREx orbital period is visible throughout the survey.
An example six-hour interval early in the survey is presented in Figure~\ref{fig:orbital}, together with the spacecraft terrestrial latitude and the corresponding SZA derived from the spacecraft telemetry. 
Smaller absolute SZA values correspond to pointing closer to the Sun, while the sign of the SZA reflects whether the spacecraft is on the sunrise or sunset half of the orbit.
The two orbital halves are not distinguished in this work.

The fitted airglow brightness varies on a $\sim$98-minute timescale, consistent with orbit-scale changes in the viewing geometry and the atmospheric conditions.
The apparent trends differ between helium and oxygen, with the helium brightness appearing to anti-correlate with oxygen brightness.
However, this apparent anti-correlation does not imply a direct physical relationship between the two emission.
Rather, the trends are largely driven by systematic trends of the survey plan that couple the observations in an anti-correlated manner.

\begin{figure}
    \centering
    \includegraphics[width=0.48\textwidth]{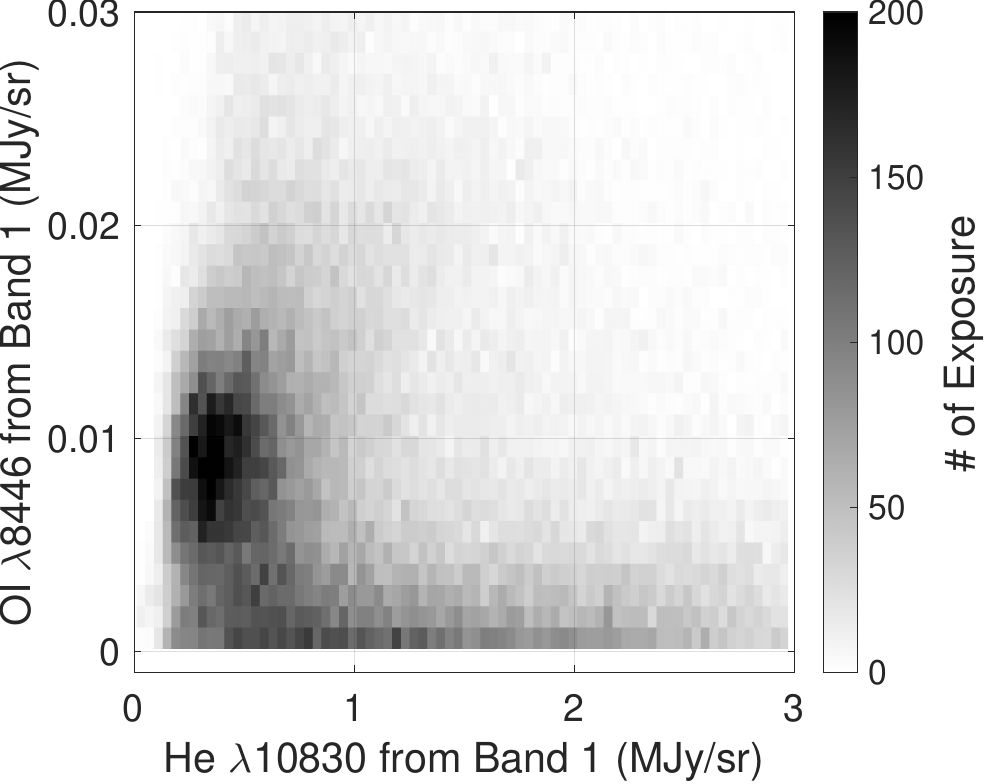}
    \caption{Two-dimensional histogram showing the relationship between \HeI\ and \OI\ brightness for individual exposures.
    The emissions show little correlation, consistent with helium and oxygen airglow being controlled by different physical mechanisms.}
    \label{fig:He_O_hist}
\end{figure}

Figure~\ref{fig:He_O_hist} compares the \HeI\ and \OI\ brightness for individual exposures.
No clear correlation between the two emissions is observed.
This indicates that the apparent anti-correlation seen in the orbital time series is not due to a direct physical relationship between helium and oxygen emission processes.
This behavior is expected because helium and oxygen airglow originate from different mechanisms in the upper atmosphere \cite{Waldrop2008,Hunten1967}.

In this example, \HeI\ varies more closely with spacecraft latitude.
Additionally, a periodic dip in brightness from $\sim$3~MJy/sr to $\sim$1~MJy/sr can be identified, coinciding with SPHEREx passing over the terrestrial South pole (Figure \ref{fig:he_trend_snapshots}).
Meanwhile, \OI\ and \OII\ rise/fall closer in phase with the proximity towards/away from the Sun respectively, as captured by the SZA.
However, the SPHEREx survey geometry couples the latitude and the SZA within each orbit, and the two parameters are not independently sampled.
Therefore, their respective influences cannot be cleanly separated using the present data alone.
Using only SPHEREx measurements, we cannot rule out an SZA dependence in helium, nor can we exclude latitude-wise structure in the oxygen emission.
Consequently, the interpretation of our results can benefit from a multivariate model accounting for both SZA and latitude dependence simultaneously.
Such modeling effort is beyond the scope of this paper.
In the next section, we present an example to alleviate the degeneracy through a subset analysis designed to isolate trends that remain coherent under different observing geometries.

\begin{figure*}[htb]
	\centering
    \includegraphics[width=0.32\linewidth]{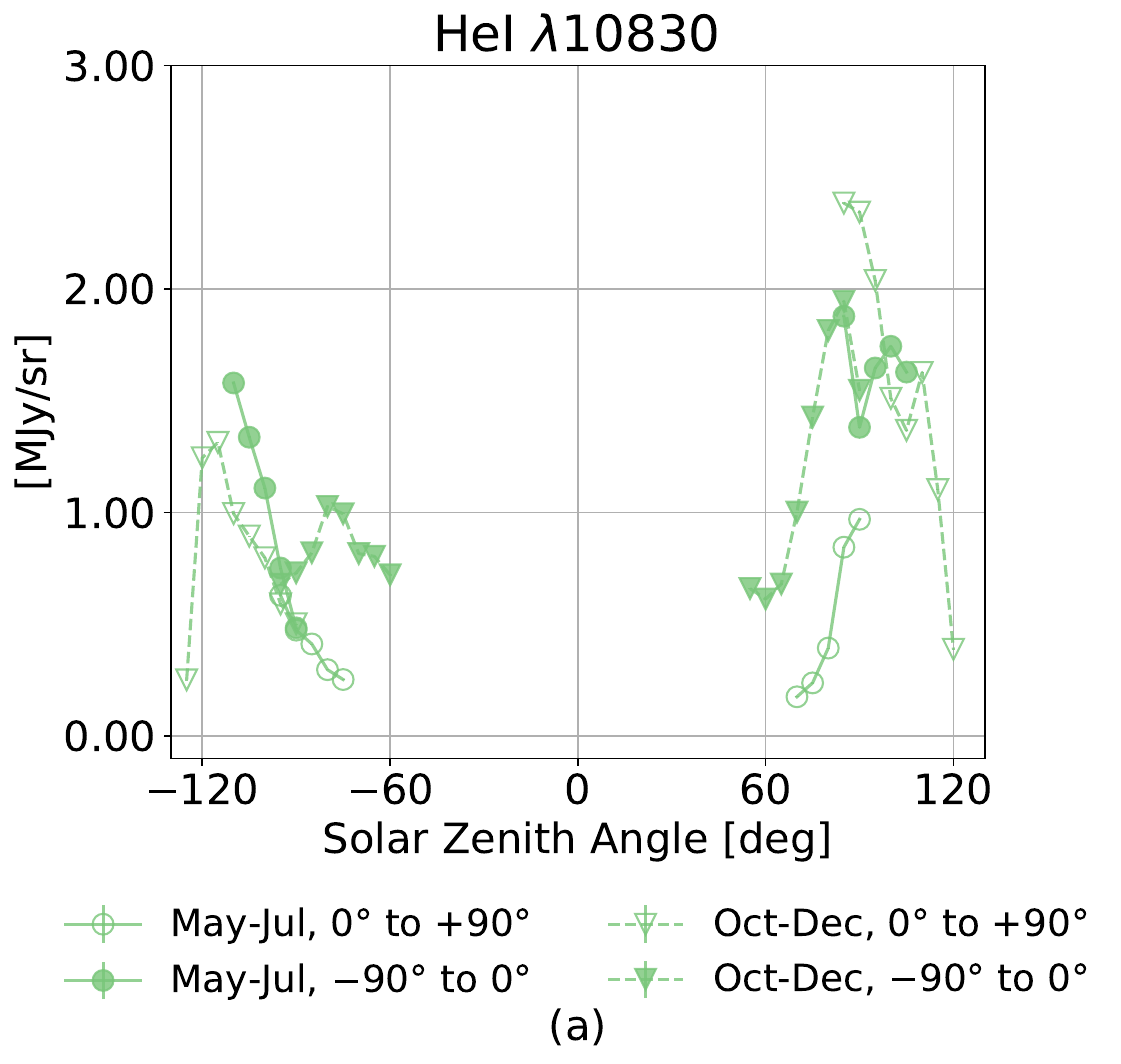}
    \hfill
    \includegraphics[width=0.32\linewidth]{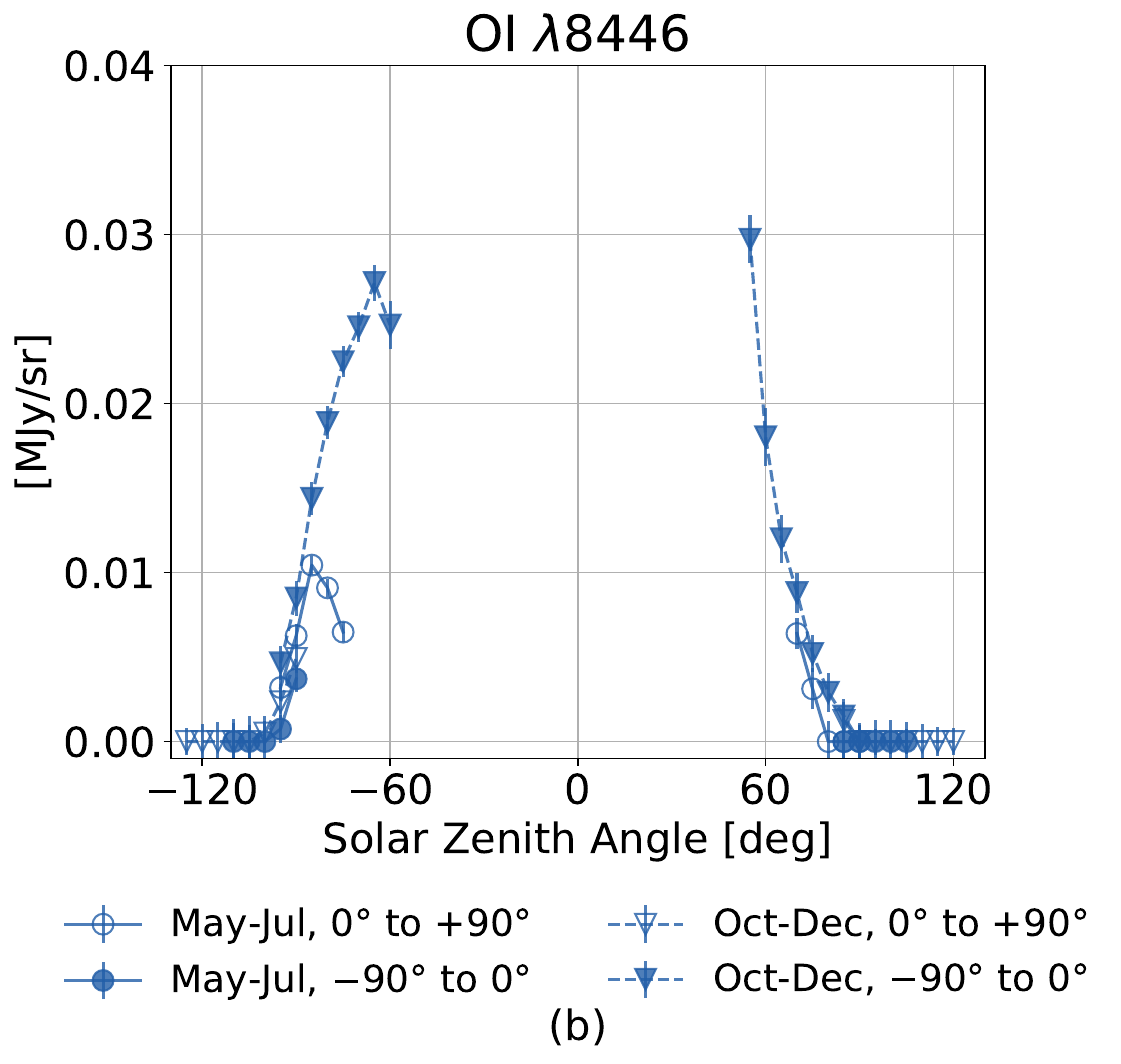}
    \hfill
    \includegraphics[width=0.32\linewidth]{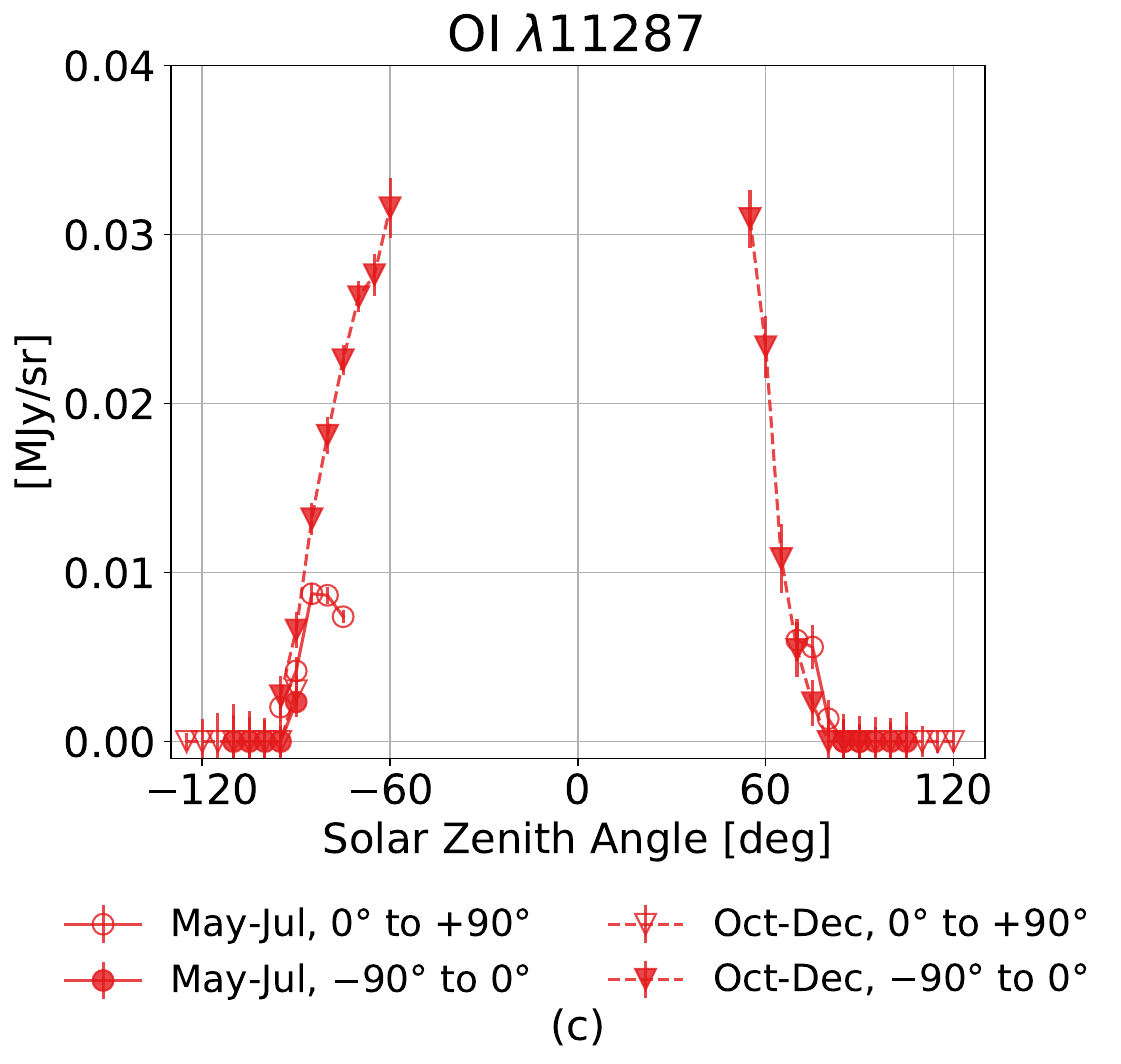}
	\caption{
    Brightness of (a) \HeI, (b) \OI, and (c) \OII\ emission as a function of the SZA in four latitude--time subsets.
    Each subset corresponds to a combination of hemisphere and three-month time interval.
    The oxygen lines show a similar dependence on SZA across all subsets, while the helium brightness exhibits larger offsets between subsets at fixed SZA.
	\label{fig:trend_zenith_lat}}
\end{figure*}

Latitude and SZA effects are partially separated by dividing the data into four subsets defined by observing timestamp and the terrestrial hemisphere.
The survey is separated into two three-month intervals, May to July and October to December.
Data from August and September are excluded because the survey was preferentially pointed toward the Galactic center during that period, limiting the sky coverage suitable for trend analysis.
Within each time interval, the data are split into the Southern ($-90^{\circ}$ to $0^{\circ}$) and the Northern ($0^{\circ}$ to $+90^{\circ}$) latitude bins.
In each latitude--time subset, the data are binned and averaged by SZA in steps of $5^{\circ}$.
The resulting line brightness as a function of SZA is plotted in Figure~\ref{fig:trend_zenith_lat}.

The \OI\ and \OII\ emission lines exhibit a coherent dependence on Solar zenith angle across all latitude--time subsets.
Within each latitude bin, the observed oxygen brightness increases toward smaller absolute SZA, consistent with enhanced excitation under stronger Solar illumination and with the atmospheric excitation framework described in Section~\ref{S:line_overview}.
The similarity of the SZA trend across all subsets suggests that oxygen emission is primarily driven by illumination geometry.

In contrast, larger dispersion between subsets at fixed SZA is seen in the helium brightness.
Although the amplitude of \HeI\ is not randomly distributed with respect to SZA, the offsets between latitude--time subsets indicate that additional spatial or seasonal structure contributes to its variability beyond the illumination geometry alone.

%% Seasonal
\subsection{Seasonal variability}

The distribution of helium in the upper thermosphere is largely controlled by global thermospheric circulation rather than local chemistry.
Interhemispheric winds driven by differential solar heating produce upwelling in the summer hemisphere and downwelling in the winter hemisphere, leading to enhanced helium abundance in the winter hemisphere \citeA{Sutton2015}.

\begin{figure*}[htb]
	\centering
	\includegraphics[width=0.48\linewidth]{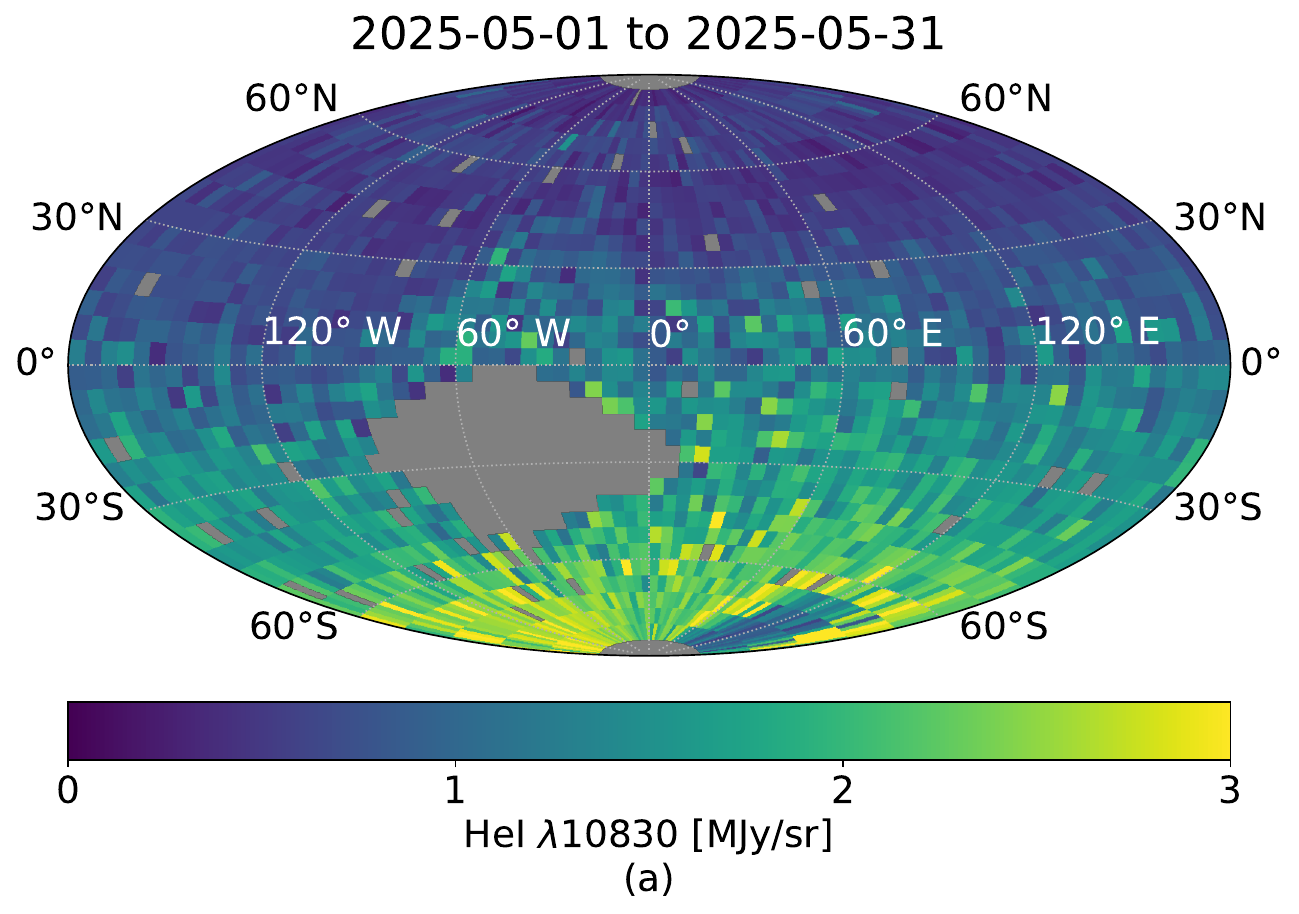}
    \includegraphics[width=0.48\linewidth]{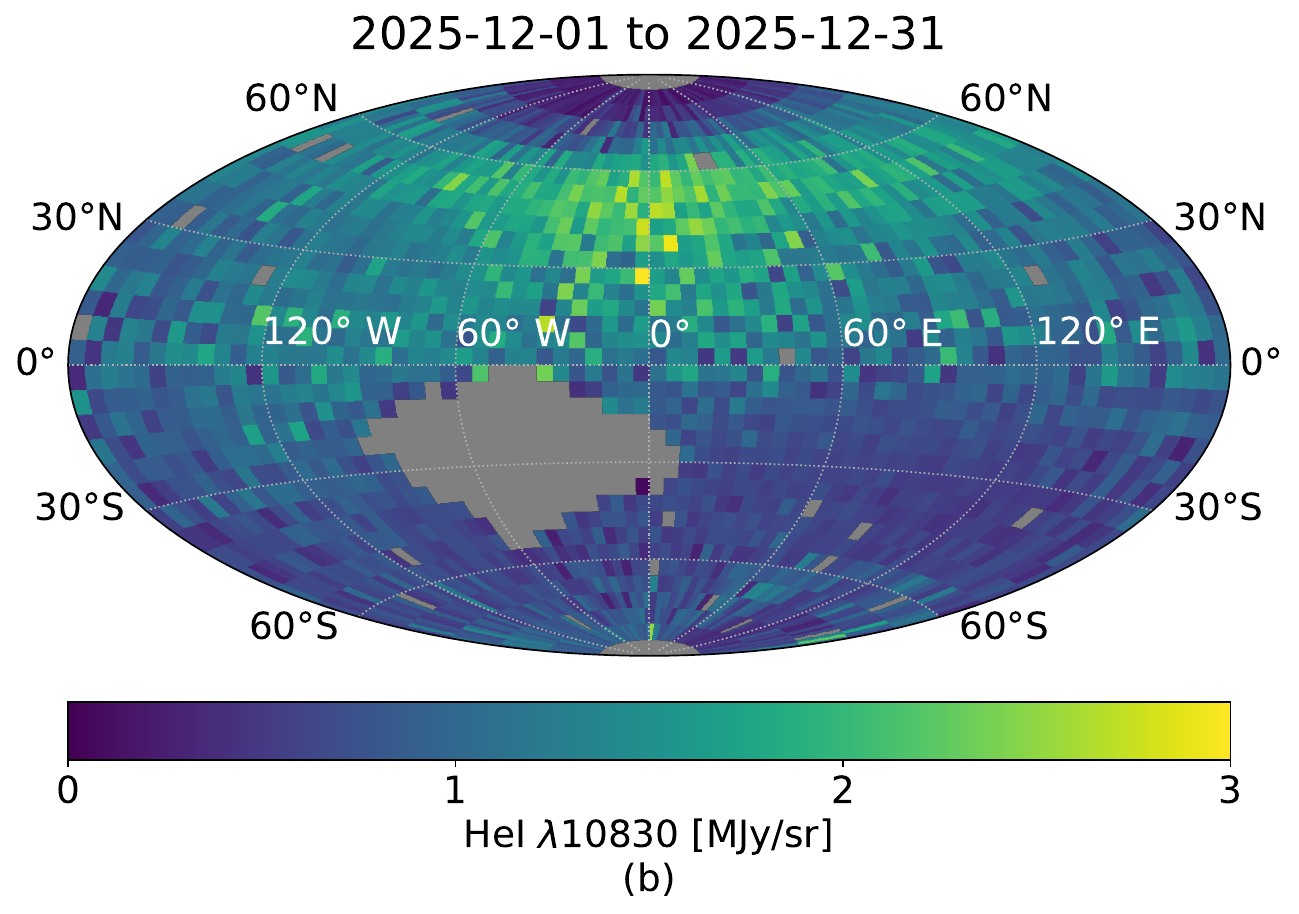}
	\caption{
	One-month snapshots of the 2-D distribution in terrestrial coordinate of the helium emission in (a) May and (b) December 2025.
    The measured emission is averaged over 31 days in latitude and longitude bins of 5-degree bin width.
    The gray diamond-shaped region in each plot is the SAA, where observations are excluded from the airglow analysis due to high transient events.
    A northward shift in the concentration of the brightest helium emission can be seen between May and December.
    An animation showing the month-by-month evolution of the helium emission is available online.
\label{fig:he_trend_snapshots}}
\end{figure*}

The two-dimensional terrestrial distribution of the \HeI\ emission is illustrated in Figure~\ref{fig:he_trend_snapshots}, covering two 31-day periods in May and December 2025.
In May 2025, the brightest emission is concentrated in the Southern Hemisphere, predominantly at latitudes $\leq -30^{\circ}$.
By December, approximately six months later, the region of enhanced emission has migrated northward and is primarily located at latitudes $\geq +30^{\circ}$.

This hemispheric shift is consistent with the seasonal evolution of the helium bulge reported in \citeA{Keating1968, Liu2014}.
The migration occurs primarily in latitude, with relatively weak dependence on the longitude.

\begin{figure*}[htb]
	\centering
	\includegraphics[width=0.48\linewidth]{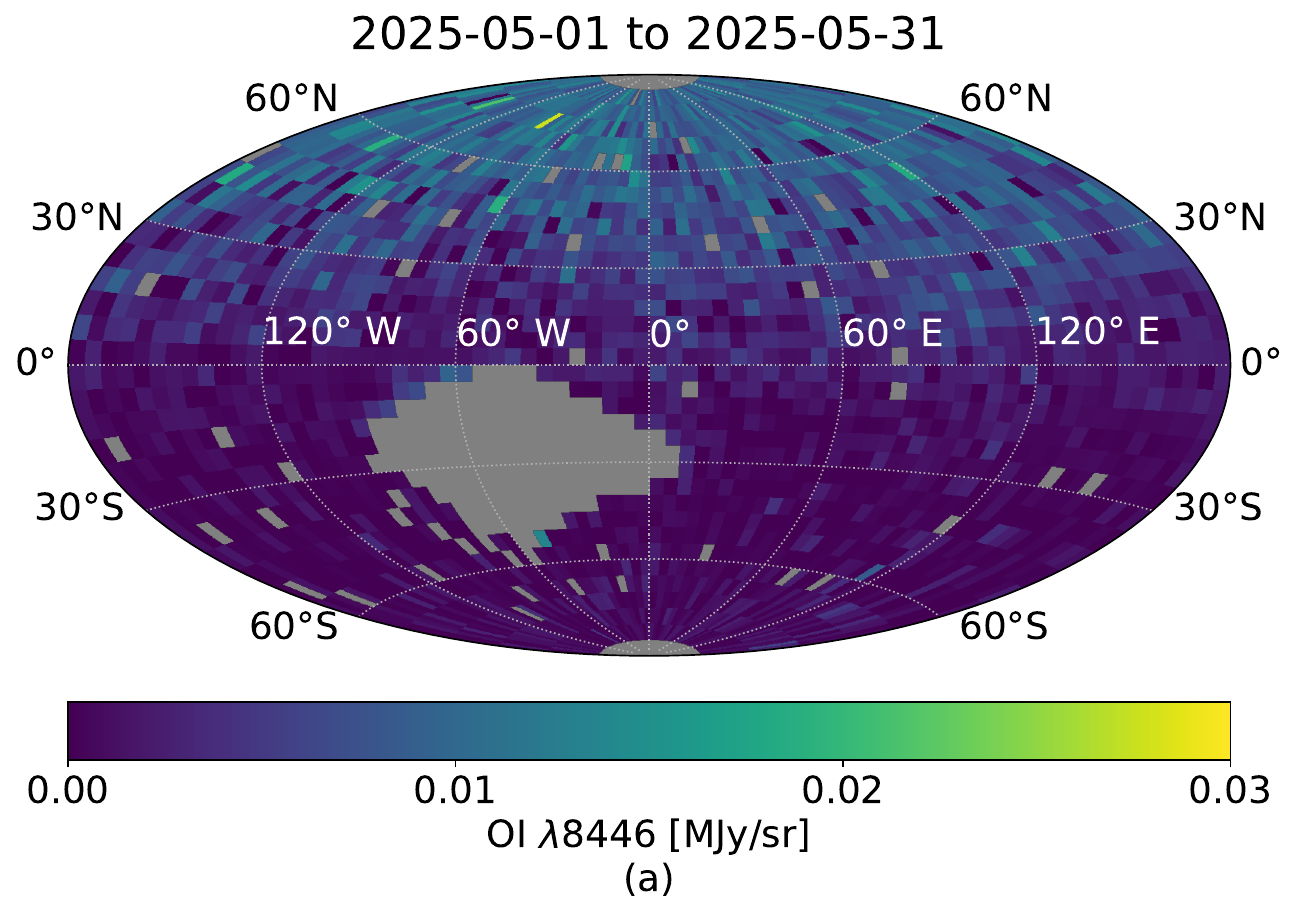}
    \hfill
    \includegraphics[width=0.48\linewidth]{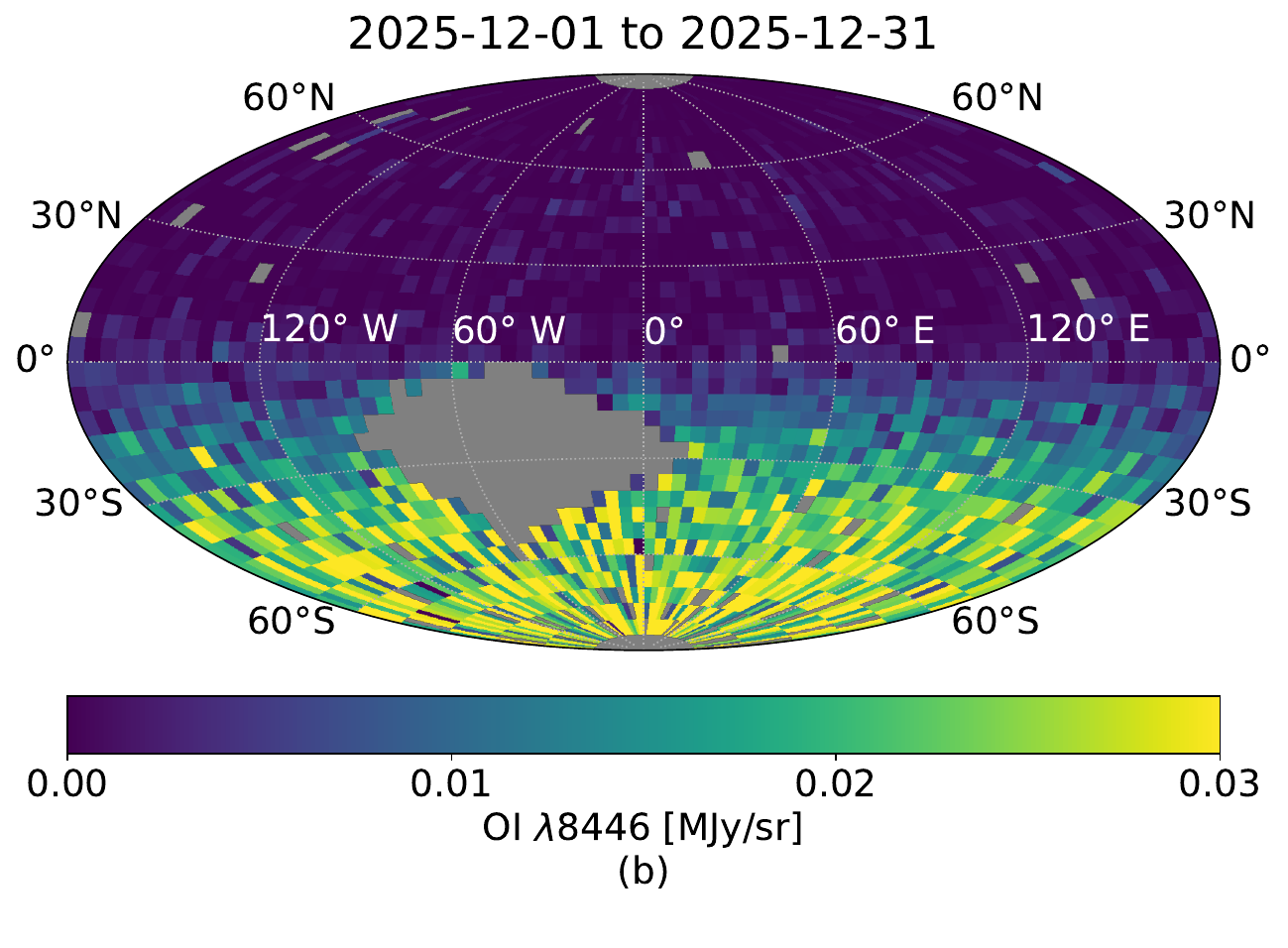}
    \hfill
    \includegraphics[width=0.48\linewidth]{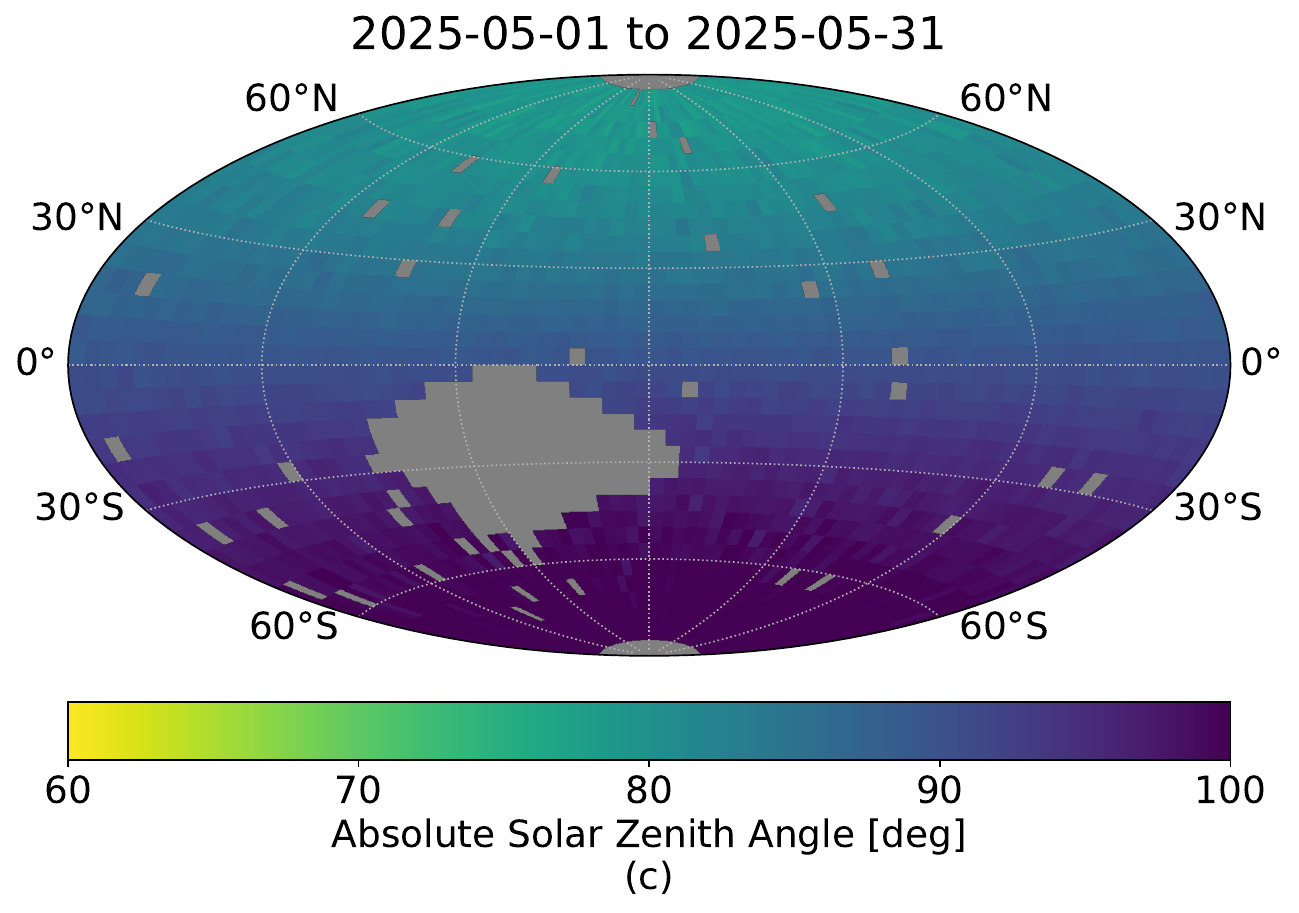}
    \hfill
    \includegraphics[width=0.48\linewidth]{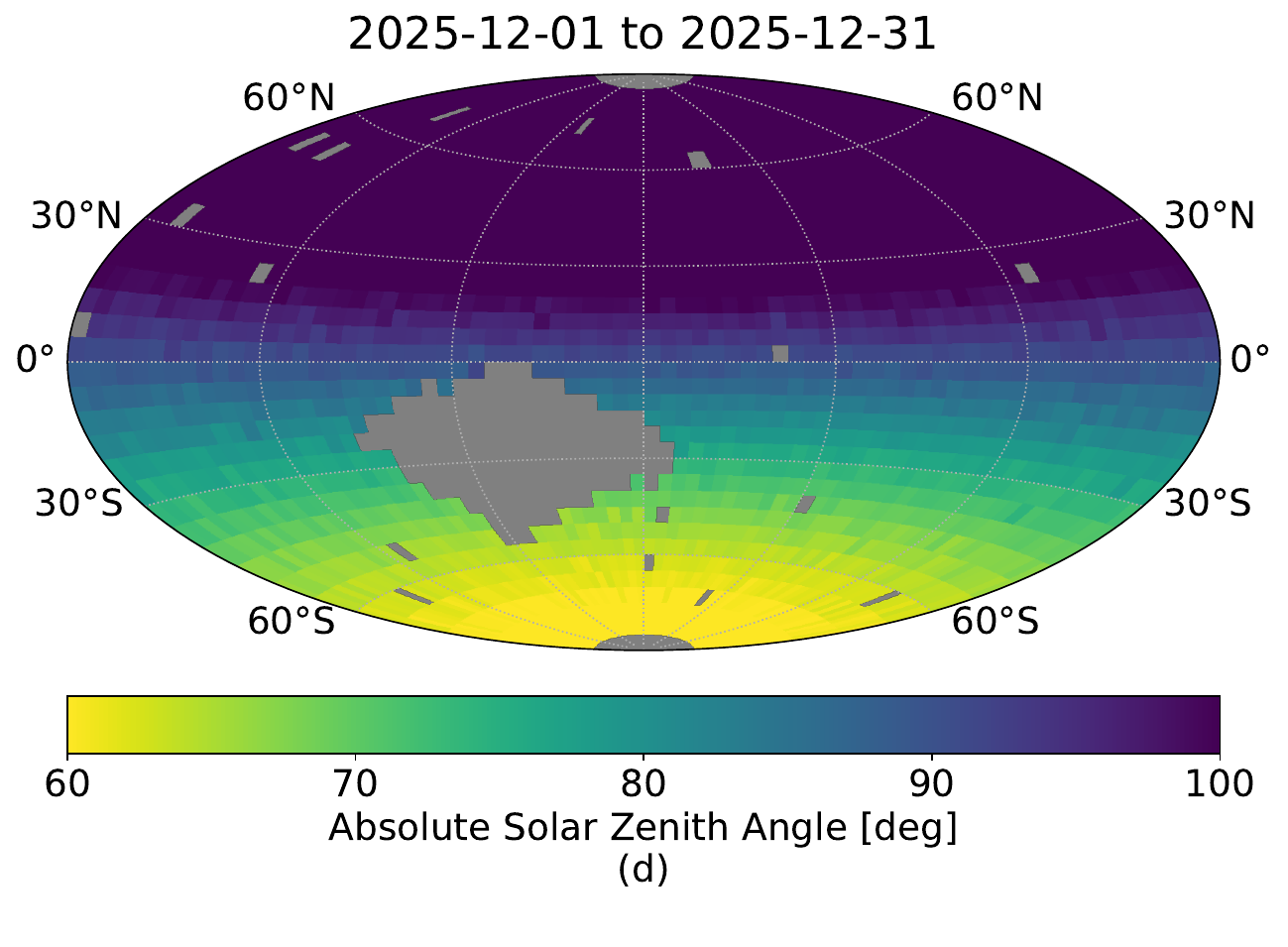}
	\caption{
	One-month snapshots of the 2-D distribution in terrestrial coordinate of the oxygen emission in (a) May and (b) December 2025.
    The measured emission is averaged over 31 days in latitude and longitude bins of 5-degree bin width.
    The corresponding absolute SZA is plotted in (c) and (d), with the color limit reversed to denote lighter color means closer pointing to the Sun.
    A movie showing the month-by-month evolution of the oxygen emission and solar zenith angle is available online.
\label{fig:oi_trend_snapshots}}
\end{figure*}

A comparable seasonal analysis for the oxygen emission is more difficult to interpret (Figure~\ref{fig:oi_trend_snapshots}).
Over long timescales, the range of sampled SZA varies substantially as the survey geometry evolves.
This behavior reflects the SPHEREx stray light avoidance constraint to optimize astronomical observational performance.
Because the oxygen brightness exhibits a strong dependence on SZA, changes in the underlying SZA distribution can mimic or obscure intrinsic seasonal structure.
Therefore, to disentangle true seasonal variation from illumination-driven effects, there is a need for a more uniform sampling of SZA than what is currently available in the present data set.

%%%% ---- SUMMARY AND FUTURE WORKS ---- %%%%
\section{Summary}
\label{S:conclusion}

This work presents global measurements of NIR atmospheric airglow in the first eight months of the SPHEREx survey, focusing on the \HeI, \OI, and \OII\ emission.
Using a multi-component spectral fitting framework that simultaneously models the ZL continuum and line templates, we extract emission-line amplitudes from more than $10^5$ exposures obtained between May and December 2025.
A signal-to-noise requirement and a goodness-of-fit cut are applied to reduce residual astrophysical contamination and to reject exposures with non-modeled broadband backgrounds.

The \HeI\ emission is detected in the majority of exposures, with a median brightness of $\sim 1~\mathrm{MJy/sr}$, exceeding the median ZL continuum near 1~\micron\ by roughly an order of magnitude.
The oxygen lines are significantly fainter, with median brightness of $\sim 0.01~\mathrm{MJy/sr}$, but are still detected at high significance in more than one third of the exposures.
When oxygen is detected, \OI\ and \OII\ are measured in the same exposures with comparable amplitudes under quiet geomagnetic conditions ($\mathrm{Kp} < 3$), providing an internal consistency check on the oxygen extraction.

The fitted amplitudes exhibit variability on multiple timescales.
During periods of highly favorable aurora condition ($\mathrm{Kp} \gg 3$), all three lines show enhanced brightness, with the largest changes at high latitudes, consistent with increased energy deposition and particle precipitation during geomagnetically active intervals.
On orbital timescales, the line amplitudes show a modulation on a cadence matching the SPHEREx orbital period.
Because the survey geometry couples the spacecraft latitude and SZA, we use a subset analysis to assess trends that remain coherent across different observing geometries.
Across all latitude--time subsets, the oxygen emission exhibits a consistent dependence on SZA, while helium shows larger offsets between subsets at fixed SZA and a clear hemispheric seasonal evolution consistent with the helium bulge.

These results establish that SPHEREx provides global, multi-season measurements of NIR airglow from LEO.
The combination of continuous sky coverage at the terminator line, repeated temporal sampling, and broad spectral range enables simultaneous characterization of helium and oxygen emission across latitudes, seasons, and geomagnetic conditions.
The present analysis provides a quantitative framework for incorporating atmospheric airglow into SPHEREx foreground modeling and demonstrates the scientific value of the survey data for studies of large-scale variability of the upper atmosphere.

The emission line templates used in this analysis are derived from the SPHEREx spectral calibration \cite{Hui2026}.
These spectral response templates are described in the SPHEREx Explanatory Supplement \cite{Akeson2025} and will be distributed as part of the SPHEREx data products.
All SPHEREx exposures are publicly available through the NASA Infrared Science Archive (\url{https://irsa.ipac.caltech.edu/Missions/spherex.html}). 
In addition, the one year data release will include the fitted airglow amplitude for each exposure, providing a useful resource for future studies of atmospheric airglow and other time variable foreground signals over the full survey.

%%%% ---- ACKNOWLEDGMENTS ---- %%%%
\acknowledgments
The authors acknowledge many extremely helpful discussions with Olga Verkhoglyadova and Panagiotis Vergados at the Jet Propulsion Laboratory, and members of the SPHEREx science and engineering teams.

We acknowledge support from the SPHEREx project under a contract from the NASA/Goddard Space Flight Center to the California Institute of Technology.

Part of the research described in this paper was carried out at the Jet Propulsion Laboratory, California Institute of Technology, under a contract with the National Aeronautics and Space Administration (80NM0018D0004).

The authors acknowledge the Texas Advanced Computing Center (TACC) at The University of Texas at Austin for providing computational resources that have contributed to the research results reported within this paper.
The High Performance Computing resources used in this investigation were provided by funding from the JPL Enterprise IT Services division.

\bibliography{agusample}

\end{document}